\documentclass[aps,prb,amsmath,amssymb,reprint,twocolumn]{revtex4-2}
\usepackage{graphicx}
\usepackage{amsmath}
\usepackage{indentfirst}
\usepackage{float}
\usepackage{subfigure}
\usepackage{bm}
\usepackage[utf8]{inputenc}
\usepackage[T1]{fontenc}
\usepackage[export]{adjustbox}

\begin{document}
	
	\title{Magnetic flux controlled current phase relationship in double Quantum Dot Josephson junction}
	
	\author{Yiyan Wang}
	\affiliation{Key Laboratory of Artificial Structures and Quantum Control (Ministry of Education), Department of Physics and 
		Astronomy, Shanghai Jiaotong University, 800 Dongchuan Road, Shanghai 200240, China}
	
	\author{Cong Li}
	\affiliation{Key Laboratory of Artificial Structures and Quantum Control (Ministry of Education), Department of Physics and 
		Astronomy, Shanghai Jiaotong University, 800 Dongchuan Road, Shanghai 200240, China}
	
	\author{Bing Dong}
	\thanks{Author to whom correspondence should be addressed. Email:bdong@sjtu.edu.cn.}
	%\email{bdong@sjtu.edu.cn.}
	\affiliation{Key Laboratory of Artificial Structures and Quantum Control (Ministry of Education), Department of Physics and 
		Astronomy, Shanghai Jiaotong University, 800 Dongchuan Road, Shanghai 200240, China}
	
	\begin{abstract}
		In this work, we study a Josephson junction with parallel-connected quantum dots (QDs) threaded by a magnetic flux in the central region. We discretize the superconducting (SC) electrode into three discrete energy levels and modify the tunneling coefficients to construct a finite-dimensional surrogate Hamiltonian. By directly diagonalizing this Hamiltonian, we compute the physical quantities of the system. Additionally, we employ a low-energy effective model to gain deeper physical insight.  
		Our findings reveal that when only one QD exhibits Coulomb interaction, the system undergoes a phase transition between singlet and doublet states. The magnetic flux has a minor influence on the singlet state but significantly affects the doublet state. When both QDs have interactions, the system undergoes two phase transitions as the SC phase difference increases: the ground state evolves from a doublet to a singlet and finally into a triplet state at $\phi = \pi$. Increasing the magnetic flux suppresses the doublet and triplet phases, eventually stabilizing the singlet state. In this regime, enhancing the interaction strength does not induce a singlet-doublet transition but instead drives a transition between upper and lower singlet states, leading to a critical current peak as $U$ increases.  
		Finally, we examine the case where the tunneling coefficient $\Gamma$ exceeds the SC pairing potential $\Delta$. Here, doublet states dominate, and the system only exhibits a phase transition between doublet and triplet states when $\phi_B = 0$. In the presence of a magnetic flux, the three states converge, resulting in a triple point in the ($\phi$, $\phi_B$) parameter space.  
		
	\end{abstract}
	\maketitle
	\section{Introduction}
	The Josephson junction has long served as a versatile platform for exploring mesoscopic quantum phenomena. It consists of a central region coupled to two superconducting (SC) leads, and its transport properties have been extensively studied in both theory\cite{josephson_junction001,josephson_junction002,josephson_junction003,josephson_junction004,josephson_junction005,josephson_junction006,josephson_junction007,josephson_junction008,josephson_junction009} and experiment\cite{josephson_junction_experiment001,josephson_junction_experiment002,josephson_junction_experiment003,josephson_junction_experiment004,josephson_junction_experiment005} A hallmark of this system is the Josephson effect: a supercurrent can flow across the junction at zero voltage bias when a finite phase difference exists between the SC leads \cite{non_bias_current001,non_bias_current002}. One of the most intriguing features of Josephson junctions is the so-called $0$–$\pi$ transition, where the Josephson current reverses direction\cite{QPT_interaction001,QPT_interaction002,QPT_interaction003,QPT_interaction004,QPT_interaction005,QPT_interaction006,QPT_magnetic001,QPT_magnetic002,QPT_magnetic003,QPT_magnetic004}. This phenomenon, which has been observed experimentally, reflects a change in the ground state of the system. In the prototypical setup with a single quantum dot (QD) as the central region, either strong on-site Coulomb interaction \cite{QPT_interaction001,QPT_interaction002,QPT_interaction003,QPT_interaction004,QPT_interaction005,QPT_interaction006} or an applied magnetic field\cite{QPT_magnetic001,QPT_magnetic002,QPT_magnetic003,QPT_magnetic004} can drive the transition. In both cases, the mechanism is similar: the ground state changes from an even-parity configuration, supporting a positive supercurrent, to an odd-parity configuration, yielding a negative one. Equivalently, the subgap Andreev bound states (ABS) cross zero energy at the quantum phase transition (QPT) point.
	
	The single-QD Josephson junction has been thoroughly investigated and is now well understood. In the absence of interactions, approaches such as nonequilibrium Green's functions \cite{PhysRevB.61.4754,YuZhu_2001}and path-integral methods\cite{ITGF} have been applied to calculate the Josephson current and its relation to ABS. With interactions present, more sophisticated analytical tools have been developed, including the slave-boson method with non-crossing approximation (NCA) \cite{PhysRevB.61.9109}, the Hubbard–Stratonovich (HS) transformation\cite{PhysRevLett.82.2788}, and functional renormalization group (FRG) techniques\cite{PhysRevB.94.085151}. tIn addition, powerful numerical methods such as numerical renormalization group (NRG)\cite{NRG001,NRG002,parallel002}and quantum Monte Carlo (QMC) simulations\cite{QMC001,QMC002,QMC003} have provided accurate benchmarks, albeit at high computational cost.
	
	Given this extensive progress, recent attention has shifted toward Josephson junctions with double quantum dots (DQDs) as the central region, explored both theoretically \cite{series001,series002,series003,series004,series005,series006,series007,parallel001,parallel002,parallel003,parallel004,parallel005,parallel006,DQDothers001,DQDothers002} and experimentally \cite{DQDxperiment001,DQDxperiment002,DQDxperiment003,DQDxperiment004,DQDxperiment005,DQDxperiment006}. Compared to the single-QD case, DQDs host a richer set of quantum states, and the associated QPTs are no longer limited to the simple $0-\pi$ transition but depend sensitively on the ground-state structure of the dots. Three typical geometries have been studied: (1) the series configuration, where current flows sequentially through two QDs \cite{series001,series002,series003,series004,series005,series006,series007}; (2) the Fano-type configuration, where one dot is directly coupled to both SCs and the other is side-coupled\cite{fano001,fano002,fano003}; and (3) the parallel configuration, where both QDs couple directly to the SC leads/[34-39]/. For the parallel case without magnetic flux, a transition from a singlet state $|\uparrow_1\downarrow_2-\downarrow_1\uparrow_2\rangle$ to triplet $|\uparrow_1\downarrow_2+\downarrow_1\uparrow_2\rangle$, $|\uparrow_1\uparrow_2\rangle$, $|\downarrow_1\downarrow_2\rangle$ has been reported\cite{PhysRevB.94.155445}. Later studies introduced magnetic flux and Rashba spin–orbit interaction into the same setup\cite{PhysRevB.98.174504}, while NRG calculations have recently examined the role of inter-dot hopping $t_d$ \cite{PhysRevB.110.125105}. However, many of these works relied on perturbation theory limited to weak tunneling, leaving the broader parameter space less understood.
	
	In parallel, new approaches have been developed to tackle such strongly correlated junctions. A notable example is the surrogate model \cite{discretized001,discretized002,discretized003,discretized004,discretized005,discretized006}which is conceptually related to the zero-bandwidth approximation. This method discretizes the SC leads into a finite number of effective levels while incorporating high-energy quasiparticle excitations via renormalized tunneling coefficients. The resulting surrogate Hamiltonian has finite dimension and can therefore be diagonalized exactly, yielding direct access to the full spectrum of energies and eigenstates. Remarkably, the surrogate model has been shown to produce quantitatively consistent results with NRG in the single-QD case, while requiring substantially less computational effort. Since the method modifies only the SC Hamiltonian, it can be naturally extended to more complex central regions, such as DQDs with magnetic flux. 
	
	In this work, we apply the surrogate model to study a Josephson junction with parallel-coupled DQDs in the presence of magnetic flux. Specifically, we discretize the BCS Hamiltonian into a three-level effective form, which is combined with the DQD Hamiltonian and a modified tunneling term that includes flux-induced phases. The resulting surrogate Hamiltonian is represented in the many-body basis and diagonalized exactly, from which physical quantities such as entropy, parity, and Josephson current are extracted. To gain further understanding, we also analyze the results using a low-energy effective model. For the noninteracting case, we additionally employ a path-integral approach to compute the Josephson current and compare it with the surrogate-model results.

	The remainder of this paper is organized as follows. In Sec. II, we introduce the full system Hamiltonian and its surrogate representation. In Sec. III, we present the diagonalization procedure and analyze the main physical quantities, distinguishing between the cases with and without Coulomb interaction. Finally, Sec. IV summarizes our findings.
	
	\section{Model Hamiltonian and Theoretical Method}
	In this work, we consider a hybrid nanodevice consisting of two quantum dots (QDs) connected in parallel between two superconducting (SC) leads, with a magnetic flux threading the loop, as illustrated in Fig .1. The Hamiltonian of the system is given by:
	\begin{equation}
		H=H_{DQD}+H_{SC}+H_T,
	\end{equation}
	with
	\begin{equation}
		\begin{split}
			H_{DQD}=&\sum_{i\sigma}\varepsilon_{i\sigma}d^{\dag}_{i\sigma}d_{i\sigma}+\sum_{i}U_{i}n_{i\uparrow}n_{i\downarrow},
		\end{split}
	\end{equation}
	\begin{equation}
		H_{SC}=\sum_{\eta \bm{k} \sigma}\varepsilon_{\eta \bm{k} \sigma}c_{\eta \bm{k} \sigma}^{\dag}c_{\eta \bm{k} \sigma}+\sum_{\eta \bm{k}}(\Delta e^{i\phi_{\eta}}c_{\eta \bm{k} \uparrow}^{\dag}c_{\eta -\bm{k} \downarrow}^{\dag}+h.c.),
	\end{equation}
	\begin{equation}
		H_{T}=\sum_{i \eta \bm{k} \sigma}(V_{\eta i} c_{\eta \bm{k} \sigma}^{\dag}d_{i\sigma}+h.c.).
	\end{equation}
	with
	\begin{equation}
		\begin{split}
		V_{L1}=V e^{-i\frac{\phi_{B}}{4}},V_{R1}=V e^{i\frac{\phi_{B}}{4}},\\
		V_{L2}=V e^{i\frac{\phi_{B}}{4}},V_{R2}=V e^{-i\frac{\phi_{B}}{4}},
		\end{split}
	\end{equation}
	Here, $d^{\dag}_{i\sigma}$($d_{i\sigma}$) create(annihilate) an electron with spin $\sigma$ and energy $\varepsilon_{i \sigma}$ on the $i$-th QD with a repulsive on-site Coulomb interaction $U_{i}$.  Similarly, $c_{\eta \bm{k} \sigma}^{\dag}$($c_{\eta \bm{k} \sigma}$) create(annihilate) an electron with spin $\sigma$, momentum $\bm{k}$, and energy $\varepsilon_{\eta \bm{k} \sigma}$ in the $\eta$-th SC lead with order parameter $\Delta e^{i\phi_{\eta}}$, where $\Delta$ and $\phi$ are real numbers. The SC-QD tunnel coupling is described by $H_{T}$, whose tunneling amplitudes $V$ are taken to be momentum independent and be set equal for simplicity.  Here, $\phi_B=\Phi_{B_0}/\Phi_0\pi $ and $\Phi_{B_0}$ is the Aharonov-Bohm (AB)
	flux threading through the closed loop indicated by the
	purple dashed lines in Fig .1. We summing the momentum summation by using the wide band approximation. Assuming the density of states as a constant, $\rho_{F} = \frac{1}{2D}$, where $D$ is half band width and set $\Gamma=2\pi\rho_{F} V^2$. Throughout we will use natural units $\hbar$ = $k_B$ = $e$ = 1.
	\begin{figure}[H]
		\centering
		\includegraphics[scale=0.08]{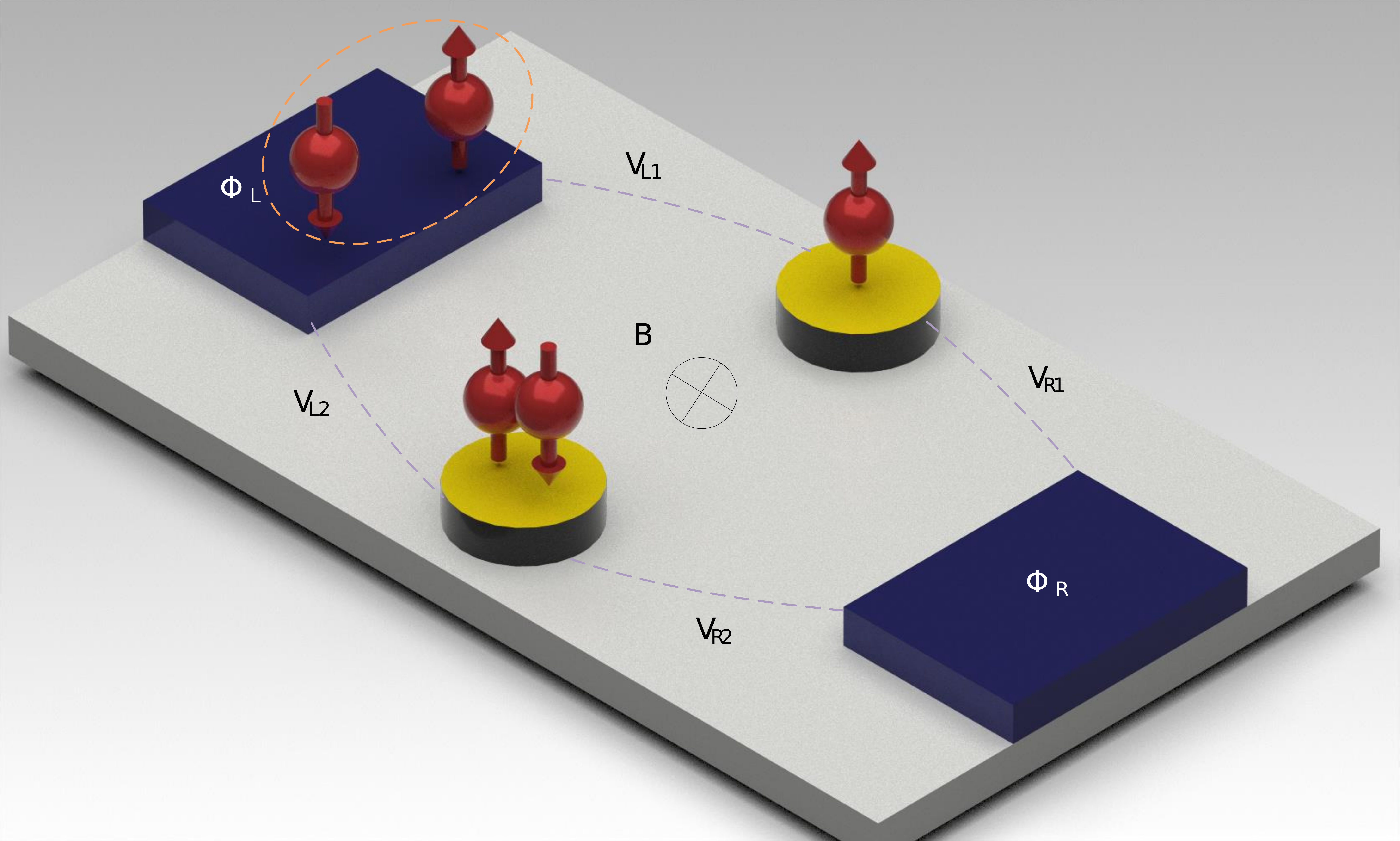}
		\caption{(Color online) Schematic diagram of two QD parallel connected to two superconductor.}
	\end{figure}
	\subsection{Surrogate SC Lead}
	We use the Nambu representation, then the tunneling self-energy is\cite{self_energy001,self_energy002}:
	\begin{equation}
		\Sigma_{T}(\omega_{n}) = -\frac{\Gamma}{2}\left[
		\begin{matrix}
			i\omega_{n} & \Delta e^{i\phi_{\eta}}\\
			\Delta e^{-i\phi_{\eta}} & i\omega_{n}
		\end{matrix}
		\right]g(\omega_{n}),
	\end{equation}
	with Matsubara frequencies $\omega_{n}=(2n+1)\pi T$, where $T$ is temperature, and the $g$ defined as
	\begin{equation}
		g(\omega) = \frac{2}{\pi}\frac{\arctan(\frac{D}{\sqrt{\Delta^2+\omega^2}})}{\sqrt{\Delta^2+\omega^2}}.
	\end{equation}
	
	According to this previous work \cite{discretized006}, here we discretize the SC lead as three points by polynomial fitting to the shape factor $g(\omega)$. Then the summation of the polynomial is $\tilde{g}(\omega)$ as:
	\begin{equation}
		\tilde{g}(\omega) = \sum_{l=-1}^{1}\frac{\gamma_{l}}{\xi_{l}^2+\Delta^2+\omega^2}.
	\end{equation}
	
	The $\tilde{g}$ function is obtained by integrating out an effective superconducting bath with the same gap $\Delta$ as the original one and whose three discrete levels with energies $\xi_{0}$ and $\pm|\xi_{1}|$ are coupled to each dot via a tunneling matrix element $\tilde{V_{l}}=\sqrt{\gamma_{l}\Gamma/2}$. Note that three is odd, so $\xi_{0}=0$. The effective bath is thus defined by parameters $\{{\tilde{\gamma}_{0}},{\tilde{\gamma}_{1}},\tilde{\xi}_{0},\tilde{\xi}_{1}\}$. And it's determined by minimizing the cost function $\chi^2 = \sum_{j}|g(\omega_{j})-\tilde{g}(\omega_{j})|^2$, which is evaluated on a non-uniformly spaced grid of frequencies.  a grid of 1000 points are logarithmically spaced in the interval $\omega \in [10^{-3}\Delta,\omega_{c}]$, and the cutting frequency is set to be $\omega_{c} = 10\Delta$ in all
	the cases analyzed below (with $D = 100\Delta$).
	
	After that, the parameters $\{{\tilde{\gamma}_{0}},{\tilde{\gamma}_{1}},\tilde{\xi}_{0},\tilde{\xi}_{1}\}$ define a surrogate Hamiltonian which replace the continuous momentum $\bm{k}$ in the raw with the discrete one $l$($l$ is an integer), and with $\xi_{\bm{k}} \rightarrow \xi_{l}$ and $V \rightarrow \tilde{V}_{l}$. We used the Mathematica "NonlinearModelFit" function to deal with the parameters fitting. The fitting residuals reach its maximum 0.015072 at $\omega=10\Delta$,which is the cutting frequency, with overall RSquared 0.99988. Then the Hamiltonian became:
	\begin{equation}
		H=H_{DQD}+\tilde{H}_{SC}+\tilde{H}_T,
		\label{SMH1}
	\end{equation}
	with
	\begin{equation}
		\tilde{H}_{SC}=\sum_{\eta l \sigma}\tilde{\xi}_{l}c_{\eta l \sigma}^{\dag}c_{\eta l \sigma}+\sum_{\eta l}(\Delta e^{i\phi_{\eta}}c_{\eta l \uparrow}^{\dag}c_{\eta l \downarrow}^{\dag}+h.c.),
		\label{SMH2}
	\end{equation}
	\begin{equation}
		\tilde{H}_{T}=\sum_{\eta i l \sigma}(\tilde{V}_{\eta i l} c_{\eta l \sigma}^{\dag}d_{i\sigma}+h.c.),
		\label{SMH3}
	\end{equation}
	where $c_{\eta l \sigma}^{\dag}$($c_{\eta l \sigma}$) create (annihilate) an electron with spin $\sigma$ and energy $\xi_{l}$ in the $\eta$-th SC lead with order parameter $\Delta e^{i\phi_{\eta}}$, and The SC-QD tunnelling amplitudes is $\tilde{V}_{\eta i l}=\tilde{V}_{l}*e^{i(\phi_B)_{\eta i}}$. Since the dimension of Hilbert space for this surrogate Hamiltonian is $4^8=65536$, it can be treated numerically and we calculated the spectrum and Josephson current by directly diagonalize the representation matrix.
	\subsection{path integral at $U_i=0$}
	If the interaction strength vanishes, the free energy and the Josephson current can be evaluated using the functional integral formalism.\cite{ITGF}
	Integrating out the leads yields the partition function:
	\begin{equation}
		Z=\int \mathcal{D} \bar{d}~\mathcal{D} d ~e^{-S_{\text{eff}}},
	\end{equation}
	with here Nambu representation
	\begin{equation}
		d=
		\begin{pmatrix}
			d_{1\uparrow}&
			d^\dagger_{1\downarrow}&
			d_{2\uparrow}&
			d_{2\downarrow}^\dagger
		\end{pmatrix}^\top,
	\end{equation}
	and 
	\begin{equation}
		S_{\text{eff}}=S_D-\int_{0}^{\beta}d\tau d\tau^\prime \bar{d}\left(\tau\right)\Sigma\left(\tau-\tau^\prime\right) d\left(\tau^\prime\right),
	\end{equation}
	where $\beta=\frac{1}{k_B T}$, and $\tau$ denotes the imaginary time. $S_D$ represents the action of the bare double dots. Performing a Fourier transform to Matsubara frequencies, the effective action $S_\text{eff}$ becomes:
	\begin{equation}
		S_\text{eff}=\sum_{\omega_{n}}\bar{d}_{\omega_{n}} \mathcal{M}_{\omega_{n}} d_{\omega_{n}},
	\end{equation}
	with $\mathcal{M}_{\omega_{n}}=-i\omega_{n}+\frac{1}{2}\left(\varepsilon_{1}+\varepsilon_2\right)\sigma_z\tau_0+\frac{1}{2}\left(\varepsilon_{1}-\varepsilon_2\right)\sigma_z\tau_z-\Sigma_{\omega_{n}}$. 
	$\sigma_i$ and $\tau_i$ denote the Pauli matrices in spin and site space, respectively. And $\Sigma_{\omega_{n}}$ is the Fourier transform of the self energy $\Sigma\left(\tau-\tau^\prime\right)$. which reads:
	\begin{equation}
		\begin{split}
			&\Sigma_{\omega_{n}}=\frac{2\Gamma}{\pi}\frac{\arctan\left(\frac{D}{\sqrt{\Delta^2+\omega_{n}^2}}\right)}{\sqrt{\Delta^2+\omega_{n}^2}}\times\\
			&\renewcommand{\arraystretch}{0.2}
			\begin{bmatrix}
				i\omega_{n}&\Delta \cos\left(\!\frac{\phi\!-\!\phi_{B}}{2}\!\right)&i\omega_{n}e^{i\frac{\phi_{B}}{2}}&\Delta \cos\left(\!\frac{\phi}{2}\!\right)\\
				\Delta \cos\left(\!\frac{\phi\!-\!\phi_{B}}{2}\!\right)&i\omega_{n}&\Delta \cos\left(\!\frac{\phi}{2}\!\right)&i\omega_{n}e^{-i\frac{\phi_{B}}{2}}\\
				i\omega_{n}e^{-i\frac{\phi_{B}}{2}}&\Delta \cos\left(\!\frac{\phi}{2}\!\right)&i\omega_{n}&\Delta \cos\left(\!\frac{\phi\!+\!\phi_{B}}{2}\!\right)\\
				\Delta \cos\left(\!\frac{\phi}{2}\!\right)&i\omega_{n}e^{i\frac{\phi_{B}}{2}}&\Delta \cos\left(\!\frac{\phi\!+\!\phi_{B}}{2}\!\right)&i\omega_{n}
			\end{bmatrix}.
		\end{split}
	\end{equation}
	Here $\phi=\phi_{L}-\phi_{R}$ denotes the phase difference between the two superconducting leads. After integrating out the Grassmann fields $\{d,\bar{d}\}$, the partition function takes the form:
	\begin{equation}
		Z=\prod_{\omega_{n}}\mathrm{det} \mathcal{M}_{\omega_{n}},
	\end{equation}
	and the Josephson current is:
	\begin{equation}
		I_J=-\frac{2}{\beta}\frac{\partial}{\partial\phi}\mathrm{ln}Z.
	\end{equation}
	\subsection{Low Energy Effective Hamiltonian}
	In the infinite-gap limit$\left(\Delta\to\infty\right)$ , quasiparticle excitations in the leads are neglected, yielding the following low-energy effective Hamiltonian:
	\begin{equation}
		\begin{split}
			H_{\text{eff}}=&H_{DQD}\\
			+\Big[&\Gamma \cos(\frac{\phi-\phi_{B}}{2})d_{1\uparrow}^{\dagger}d_{1\downarrow}^{\dagger}+\Gamma \cos(\frac{\phi}{2})d_{1\uparrow}^{\dagger}d_{2\downarrow}^{\dagger}\\
			+&\Gamma \cos(\frac{\phi+\phi_{B}}{2})d_{2\uparrow}^{\dagger}d_{2\downarrow}^{\dagger}+\Gamma \cos(\frac{\phi}{2})d_{2\uparrow}^{\dagger}d_{1\downarrow}^{\dagger}+h.c.\Big].
		\end{split}
	\end{equation}
	In the presence of interactions, we directly construct the matrix representation of the Hamiltonian. The full Hilbert space is $16 \times 16$, which allows for an analytical solution.
	
	\section{Results and Discussions}
	In this section, we numerically solve the finite-dimensional surrogate Hamiltonian. The eigenvectors are then employed to evaluate various physical quantities, including entropy, parity, spin correlations, and the Josephson current. We focus on the particle-hole(ph) symmetric point$\left(\varepsilon_i \!=\!-\!\frac{U_i}{2}\right)$. We set $\phi_{L}\!=\!\frac{\phi}{2}$ and $\phi_{R}\!=\!-\!\frac{\phi}{2}$ without loss of generality. The temperature is fixed at zero for surrogate Hamiltonian, and the superconducting gap $\Delta$ is taken as the unit of energy, with $\Delta=1$. 
	\subsection{$U_1=0, U_2=0$}
	The first case we consider is the non-interacting limit. In this case, we calculate the Josephson current under varying magnetic flux using both the surrogate Hamiltonian and the path integral formalism. For the surrogate Hamiltonian, the Josephson current is computed as $I_J=2\frac{\partial F}{\partial\phi}$. Here, 
	F is the free energy, which is given by
	\begin{equation}
		F = -T\ln[\sum_{n}e^{-E_{n}/T}],
	\end{equation}
	where $E_{n}$ is the eigvalue of the surrogate Hamiltonian.
	The result are shown in Fig .2.  
	\begin{figure}[h]

		\centering
		\includegraphics[scale=0.28]{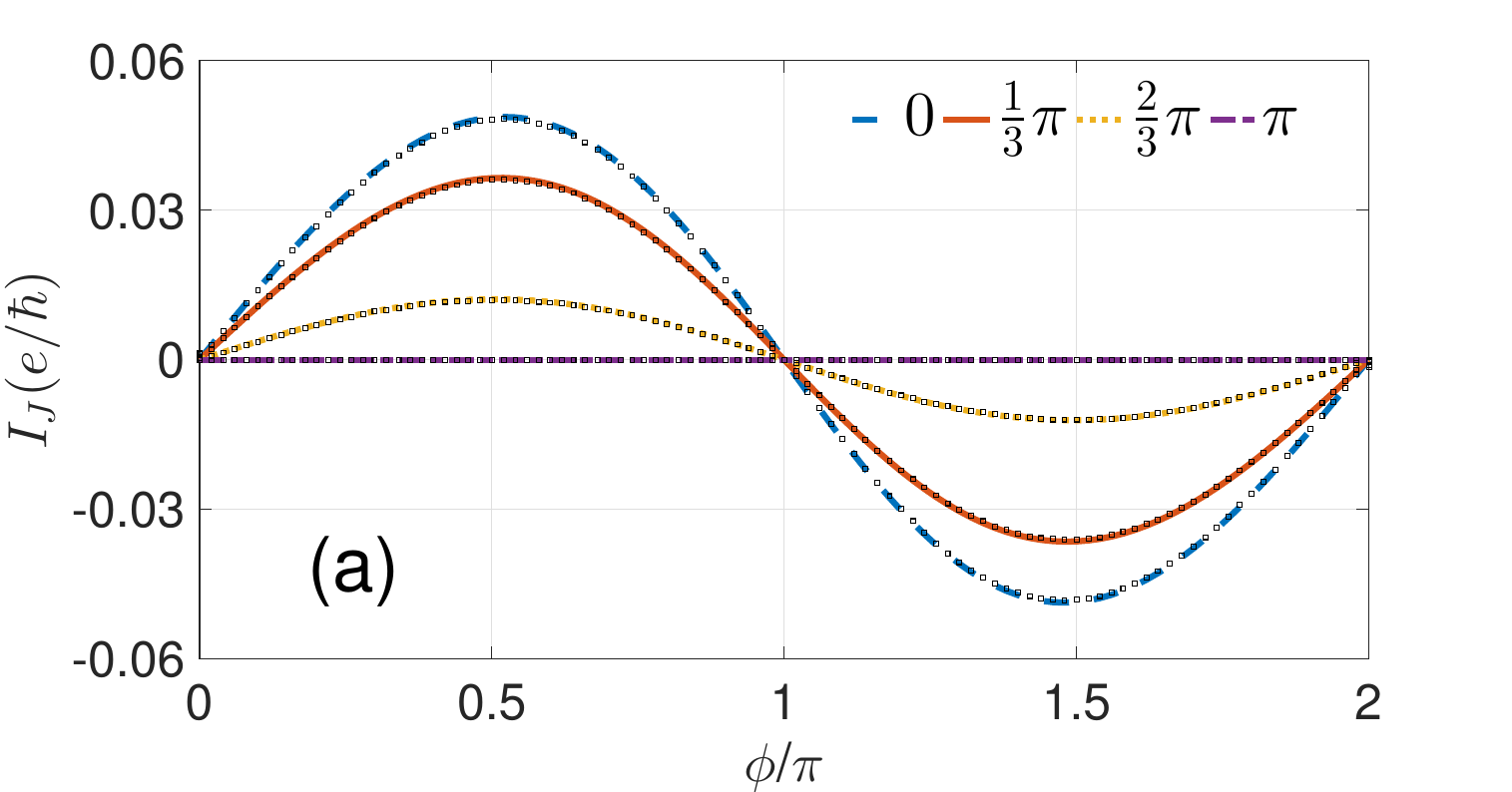}
		\includegraphics[scale=0.28]{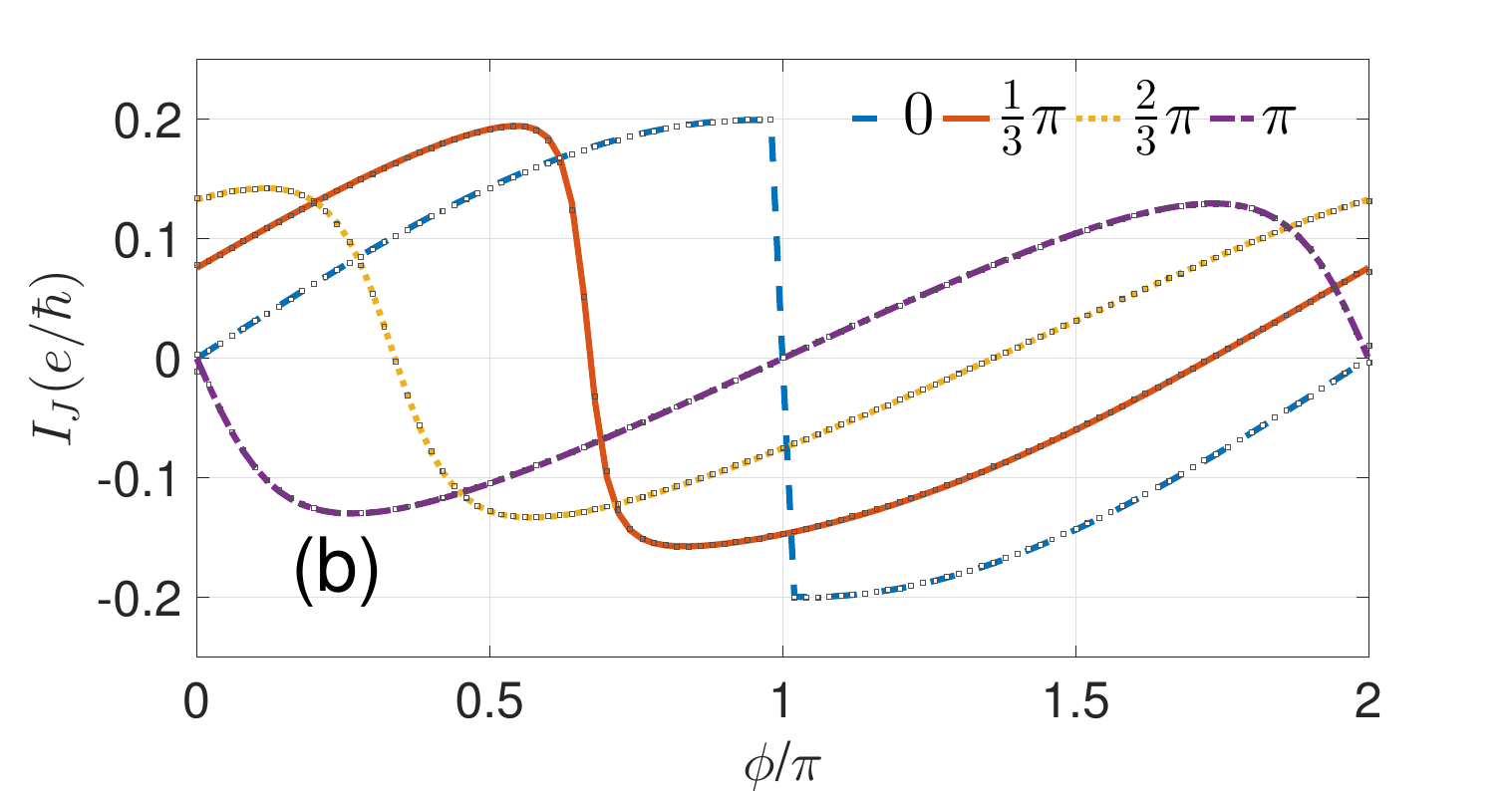}
		\includegraphics[scale=0.28]{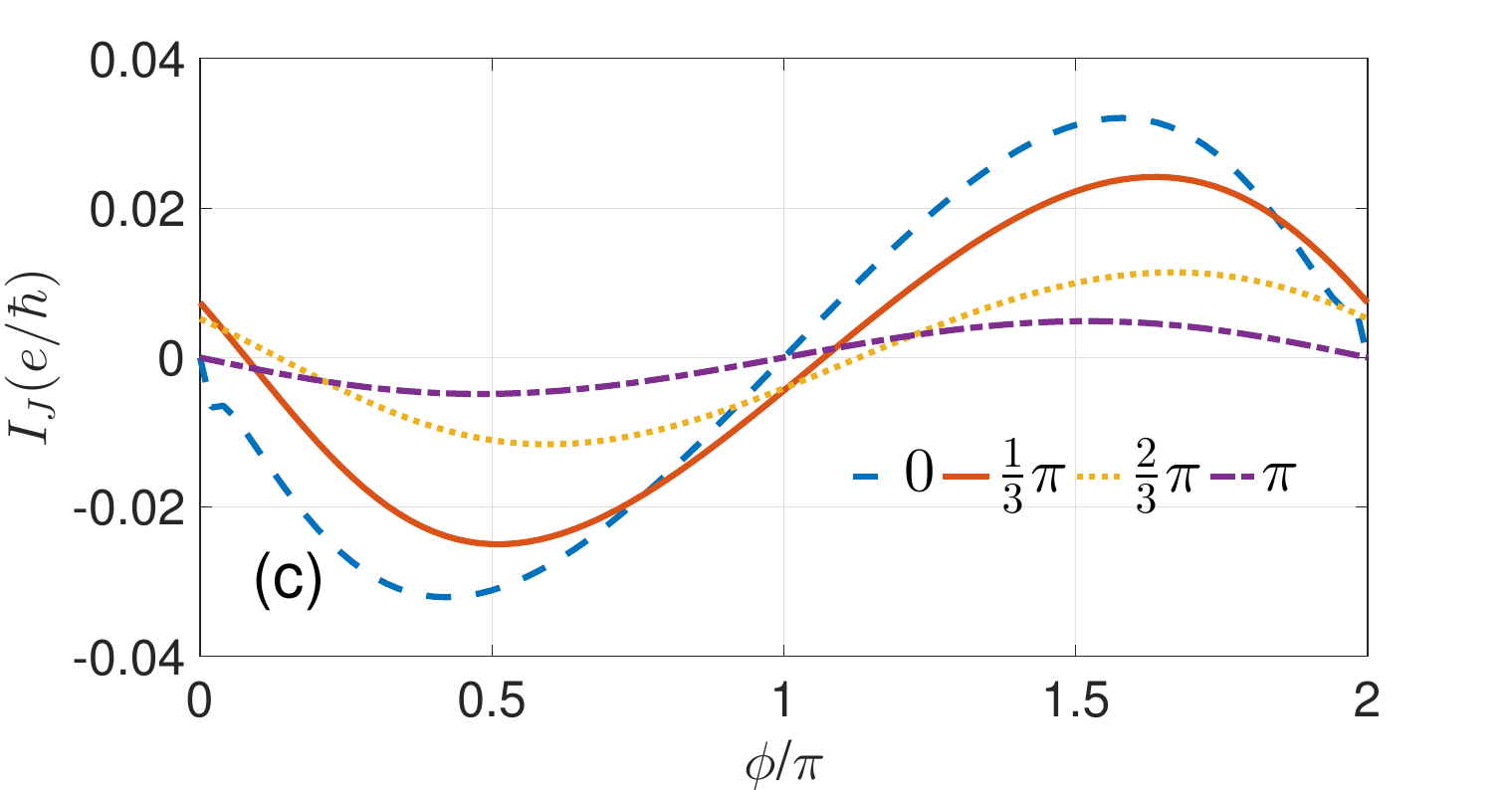}
		\includegraphics[scale=0.28]{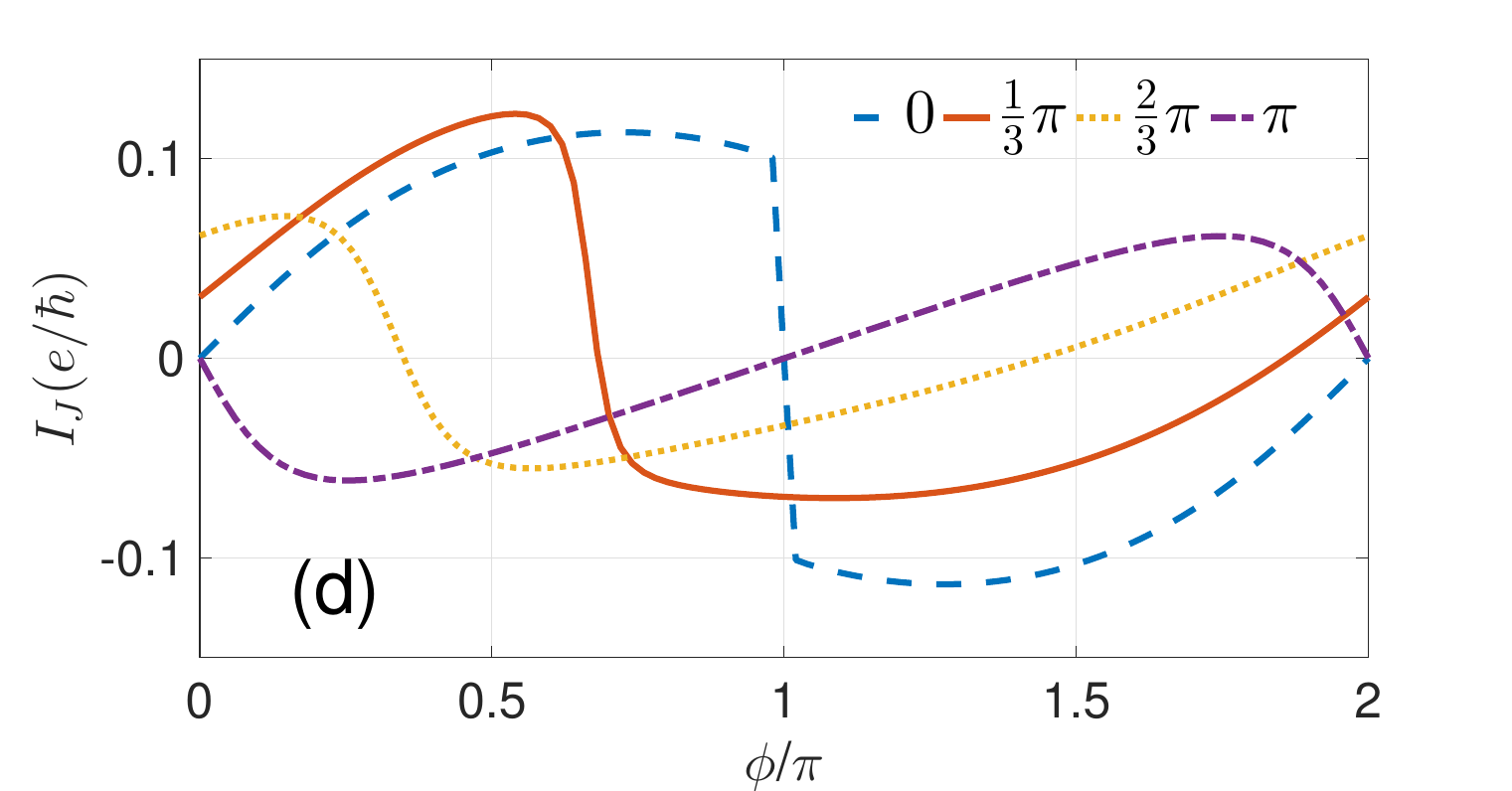}
		\caption{(Color online) (a) Current suppressed with the different $\phi_{B}$. $\varepsilon_1=\varepsilon_2=-1$ and $\Gamma=0.25$. Blank square curves are currents calculated by surrogate Hamiltonian. (b) Current is non-zero at $\phi_{B}=\pi$. $\varepsilon_1=0, \varepsilon_2=-1$ and $\Gamma=0.25$. Blank square curves are current calculated by surrogate Hamiltonian. (c) Continuum current at different $\phi_{B}$. (d) Sub-gap current at different $\phi_{B}$}
	\end{figure}
	When the dots are set symmetrically, i.e., $\varepsilon_{1}\!=\!\varepsilon_{2}\!=\!\varepsilon\!<\!0$, the Josephson current is gradually suppressed as the $\phi_{B}$ increase from 0 to $\pi$. And when the $\phi_{B}$ reaches $\pi$, the Josephson current vanish. As it is shown in Fig .2(a). This arises because:
	\begin{equation}
		\begin{split}
			&\hspace{0.5cm}Z\left(\phi_{B}=\pi\right)\\
			&=\prod_{n\in\mathbb{Z}^+}\mathrm{det}\left(\mathcal{M}_{\omega_{n+1}}\!\!\cdot\mathcal{M}_{\omega_{-n}}\right)\\
			&=\!\prod_{n\in\mathbb{Z}^+}\!\left[\left(1\!+\!\frac{\Gamma/2}{\sqrt{\Delta^2\!+\!\omega_{n-1}^2}}\right)^2\!\omega_{n-1}^2\!+\!\varepsilon^2\!+\!\frac{\Gamma^2\Delta^2}{4\left(\Delta^2\!+\!\omega_{n-1}^2\right)}\right]^4.
		\end{split}
	\end{equation}
	For simplicity, here the bandwidth D is taken to be $D\!\to\!\infty$. The partition function becomes independent of $\phi$ when $\phi_B\! =\!\pi$ and $I_J=-\frac{2}{\beta}\frac{\partial}{\partial\phi}\mathrm{ln}Z=0$
	
	If the on-site energies of the two dots are unequal($\varepsilon_{1}\neq\varepsilon_{2}$), the Josephson current remains nonzero at $\phi_{B}\!=\!\pi$, since the two tunneling paths are no longer equivalent and cannot fully cancel each other. Moreover, the current direction reverses as the phase changes from 0 to $\pi$, as shown in Fig .2(b). 
	
	By transforming  the Matsubara summation into the contour integral$(\omega_{n}\to z)$, we are able to distinguish the current contributed by Andreev bound state or by continuum quasi-particle excitation in the lead:
	\begin{equation}
		\begin{split}
			I_J&=-\frac{2}{\beta}\frac{\partial}{\partial\phi}\mathrm{ln}\left(\mathrm{det} \mathcal{M}_{\omega_{n}}\right)\\
			&=\frac{1}{2\pi i}\!\lim_{\eta\to 0^+}\!\int_{-\infty}^{-\Delta}\!\mathrm{d}z\left[\frac{\partial\!\left|\mathcal{M}\right|\!/\partial\phi}{\left|\mathcal{M}\right|(z\!+\!i\eta)}\!-\!\frac{\partial\!\left|\mathcal{M}\right|\!/\partial\phi}{\left|\mathcal{M}\right|(z\!-\!i\eta)}\right]n_f(z)\\
			&\hspace*{0.2cm}+\frac{1}{2\pi i}\!\lim_{\eta\to 0^+}\!\int_{\Delta}^{\infty}\!\mathrm{d}z\left[\frac{\partial\!\left|\mathcal{M}\right|\!/\partial\phi}{\left|\mathcal{M}\right|(z\!+\!i\eta)}\!-\!\frac{\partial\!\left|\mathcal{M}\right|\!/\partial\phi}{\left|\mathcal{M}\right|(z\!-\!i\eta)}\right]n_f(z)\\
			&\hspace*{0.2cm}-\sum_{E^\star}^{}n_f(E^\star)\operatorname{Res}_{z = E^\star} \left[ \frac{\partial\!\left|\mathcal{M}\right|\!/\partial \phi}{\left|\mathcal{M}\right|(z)}  \right],
		\end{split}
	\end{equation}
	Where $n_f(z)=\frac{1}{e^{\beta z}+1}$ which has poles at $z=i\omega_{n}$. And $E^\star$ are the Andreev bound state energy, corresponding to the isolated poles. The first two terms define the contribution of the continuum current while the last one define the discrete counterpart. Parameters are chosen as in the Fig .2(b).
	
	As it is shown in Fig .2(c) and Fig .2(d), by changing the magnetic flux, the continuum current is affected slightly, although this part contribute little, with maximum absolute value 0.03 at SC phase difference approximately $0.5\pi$ at $\phi_{B}=\pi$. The maximum value is gradually decreased to 0.01 at $\phi_{B}=\pi$. Through out, the continuum currents are almost $\pi$ phase. And the critical phase where the Josephson current reaches its maximum is varying around $0.5\pi$
	
	On the other hand, the Andreev bound state current dominates the total Josephson current and is strongly affected by the magnetic flux. The critical phases are shifted with the $\phi_{B}$ and the current reach the $\pi$ phase at $\phi_{B}=\pi$. The maximum value of the current is decreased as well as the shape of current is changed.
	
	These two contributions together constitute the total Josephson current. Unlike the phase transition in S-QD-S system\cite{QPT_interaction003}, the $0-\pi$ transition here preserves the parity, which remains 0 and the entropy is always $\mathrm{ln} 1$ as is shown in Fig .2(e) and Fig .2(f). This indicates that degeneracy is kept to be 1. This behavior arises because the maximum and minimum energy points are shifted by a phase $\pi$ due to the presence of the magnetic flux, while the different energy levels do not cross.
	\subsection{$U_1\neq 0, U_2=0$}
	In this case, the Josephson current does not vanish at $\phi_{B}=\pi$ even for a symmetric dots configuration. Here we directly diagonalized surrogate Hamiltonian to obtain numeric results. A low energy effective model is then employed to interpret these results and provide a physical insight. Initially, we set the interaction on one of the two dots to zero. The main result are shown in Fig .3.
	\begin{figure}[h]
		\centering
		\includegraphics[width=0.95\linewidth]{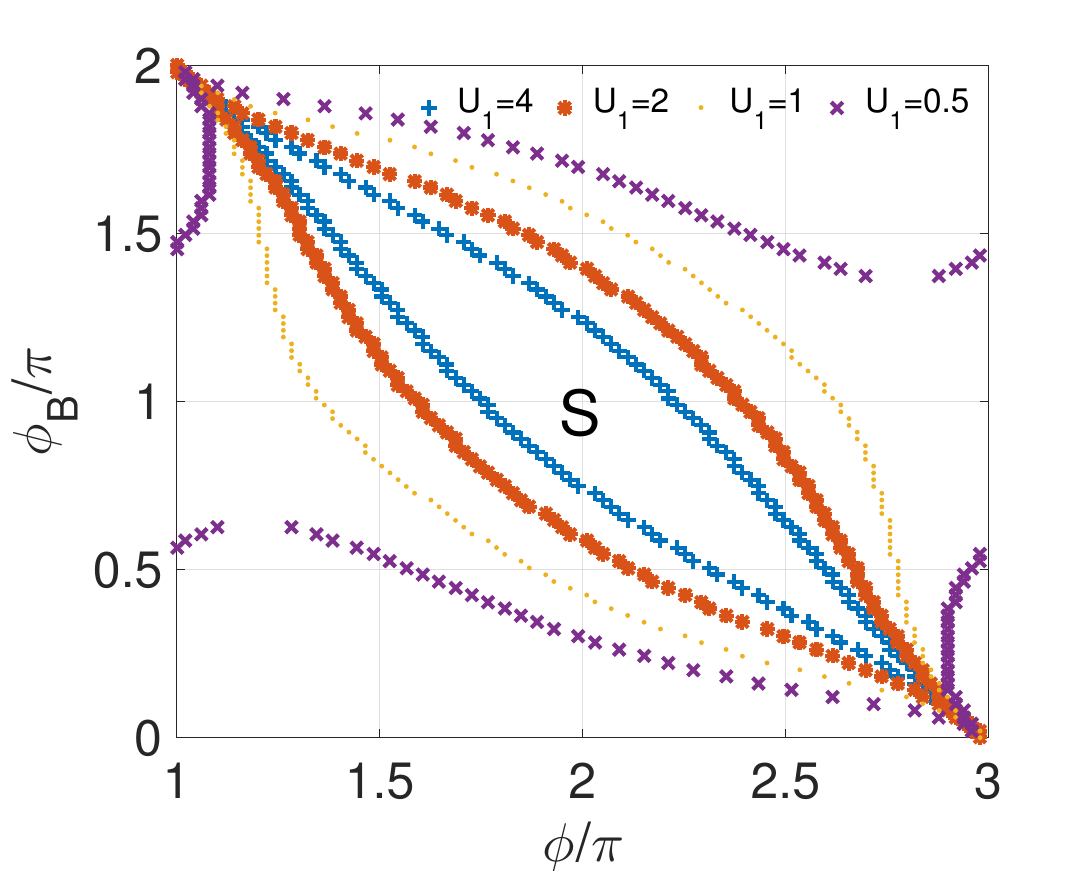}	
		\caption{(Color online)  Phase diagram as the function of $\phi$ and $\phi_{B}$. The curves denotes the phase boundary. Parameters are set as follows: $\varepsilon_1=-\frac{U_1}{2},\varepsilon_2=0$ and $\Gamma=0.25$. }
	\end{figure} 
	A phase transition occurs between the singlet $|S\rangle$ state, whose parity is 0, and the doublet state, whose parity is $\pm 1$. In Fig .3(a), we show the phase boundary:the inner region corresponds to the ground state with parity 0 $\left(|S\rangle\, state\right)$, while the outer region corresponds to the ground state with parity 1$\left(doublet\, state\right)$. It can be seen that $\phi_{B}$ significantly affects the ground-state parity. For moderate interaction strengths$(U_1\gtrsim1)$, the center of the phase boundary shifts almost linearly with $\phi_{B}$. As the interaction strength increasing, and the singlet-phase region shrinks as the interaction strength increases.
	
	For small interaction strengths, however, the singlet-phase region expands and the boundary deforms. Starting from $\phi_{B}=\pi$, it can eventually fill the entire parameter space, indicating that the ground state is a singlet, in agreement with the results shown in Fig .2(e).
	
	To interpret this behavior, we employ the low-energy effective model [Eq(19)], and the corresponding results are shown in Fig .4.
	\begin{figure}[h]
		\begin{minipage}{0.49\linewidth}
			\centering
			\includegraphics[width=1\linewidth]{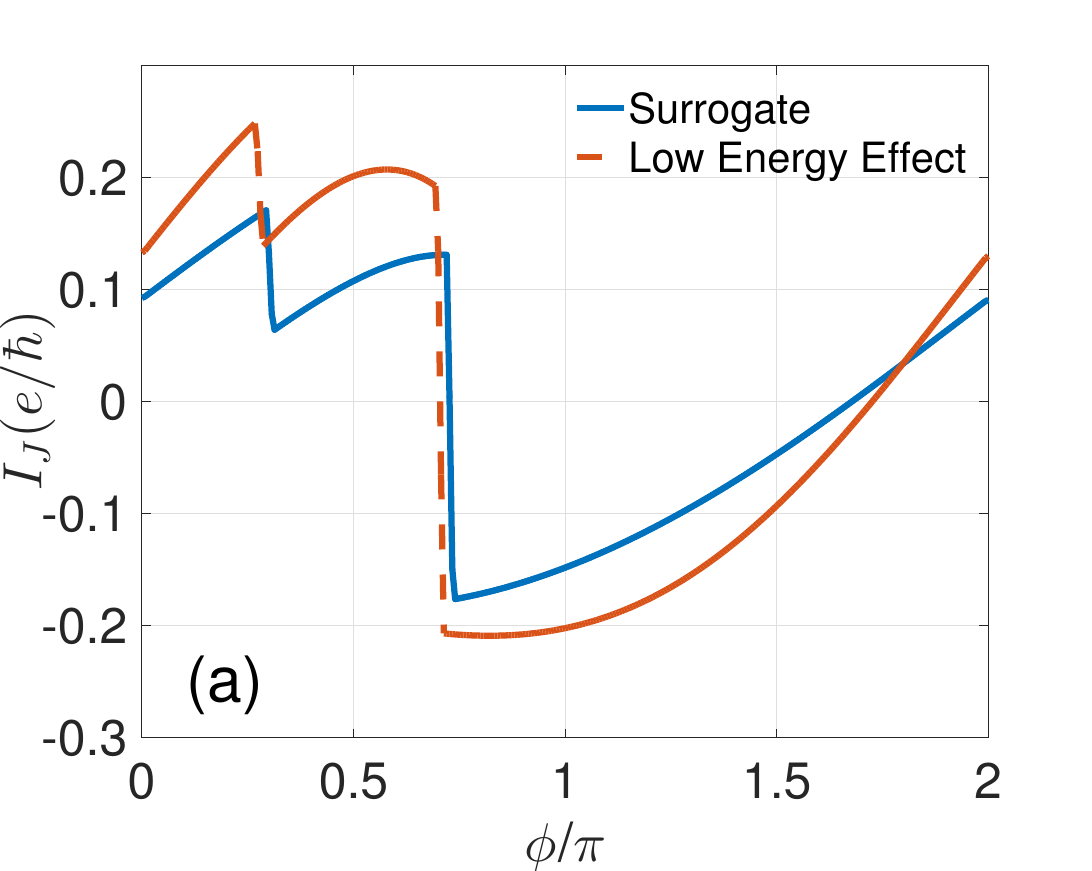}
			\includegraphics[width=1\linewidth]{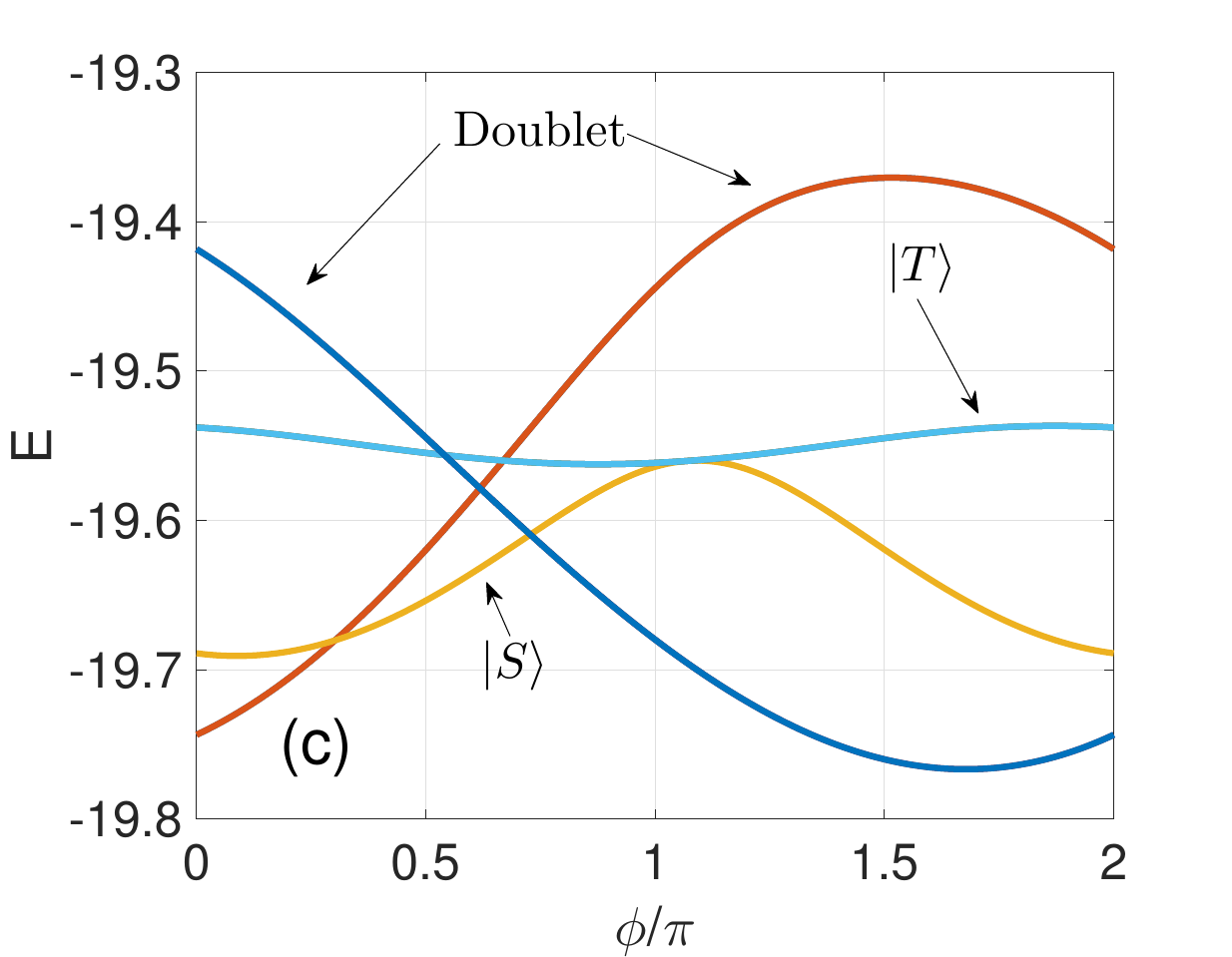}
			\includegraphics[width=1\linewidth]{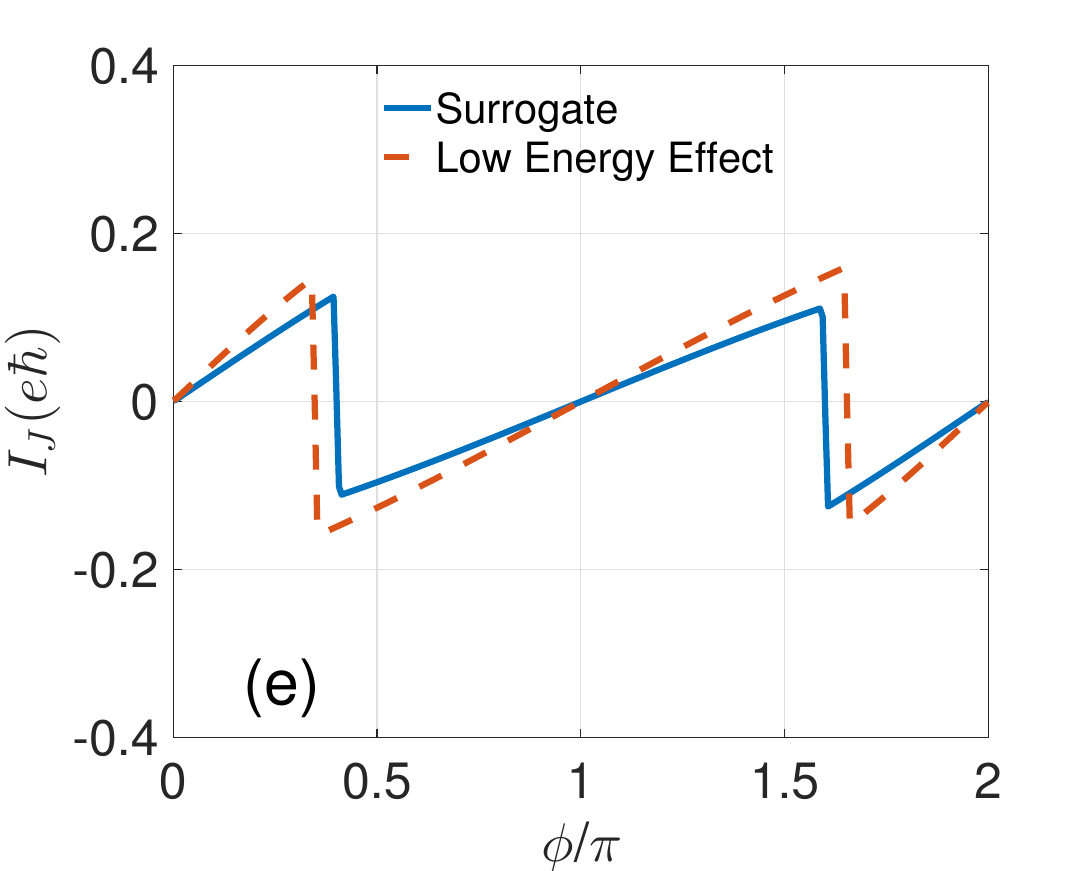}
			\includegraphics[width=1\linewidth]{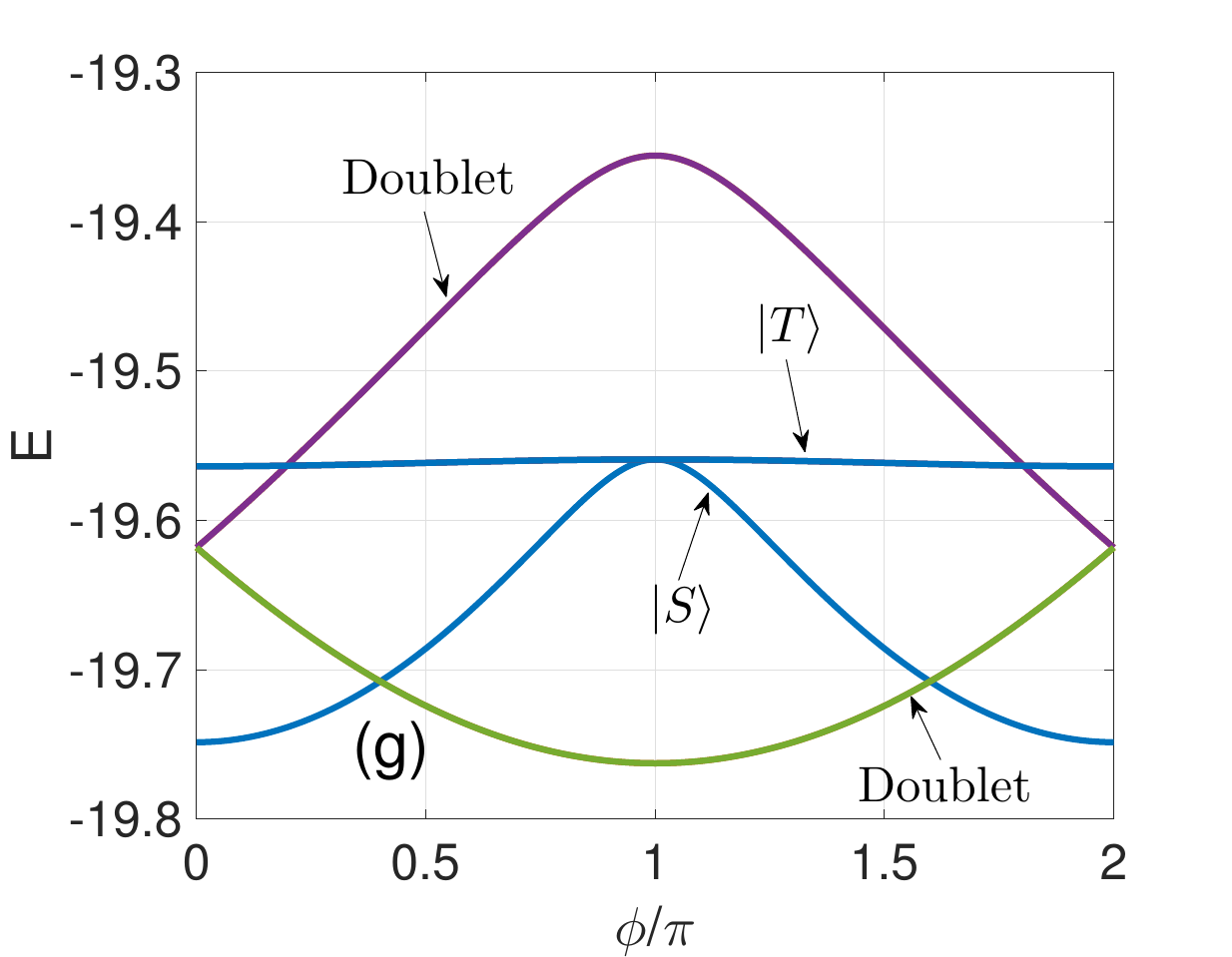}
		\end{minipage}	
		\begin{minipage}{0.49\linewidth}
			\centering
			\includegraphics[width=1\linewidth]{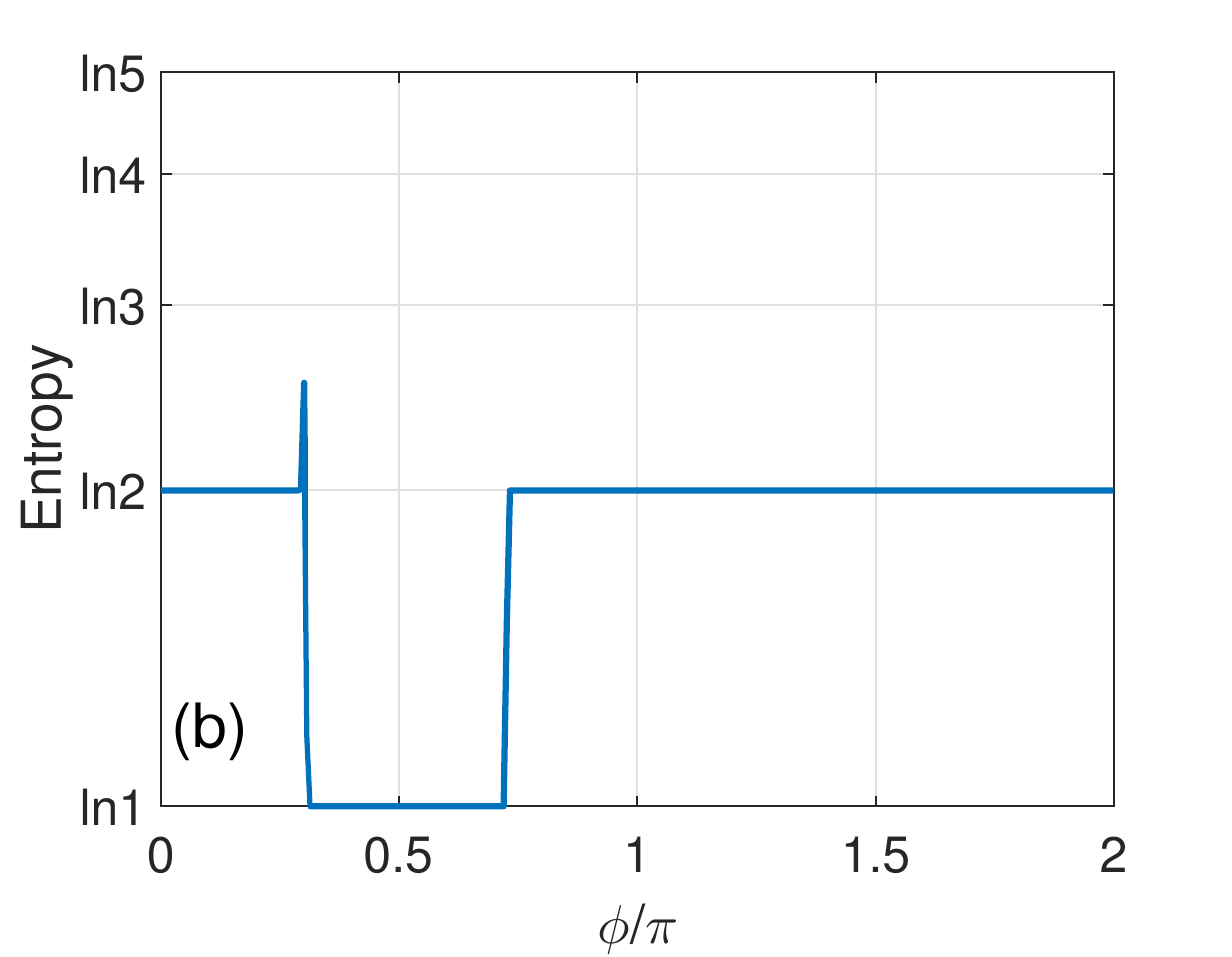}
			\includegraphics[width=1\linewidth]{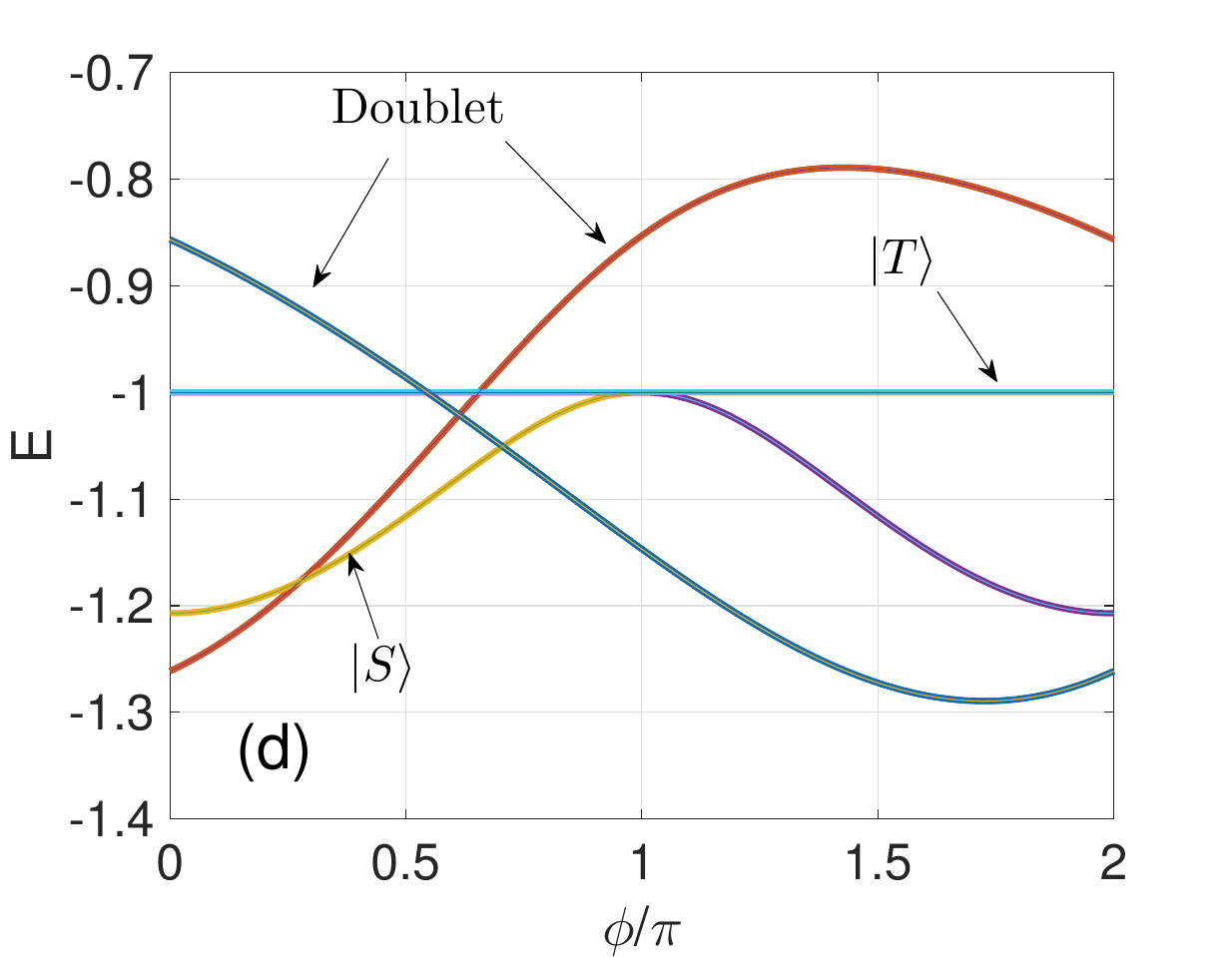}
			\includegraphics[width=1\linewidth]{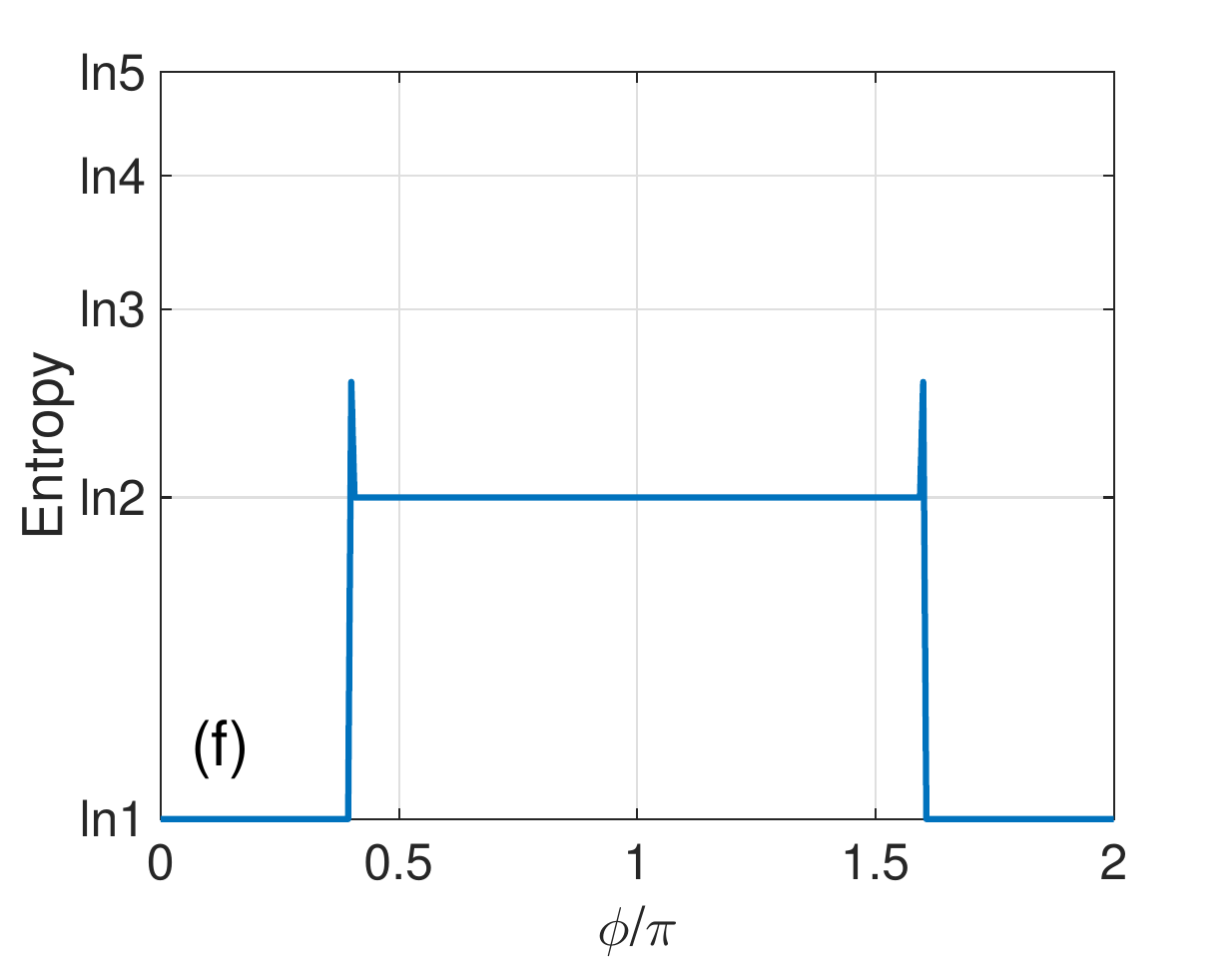}
			\includegraphics[width=1\linewidth]{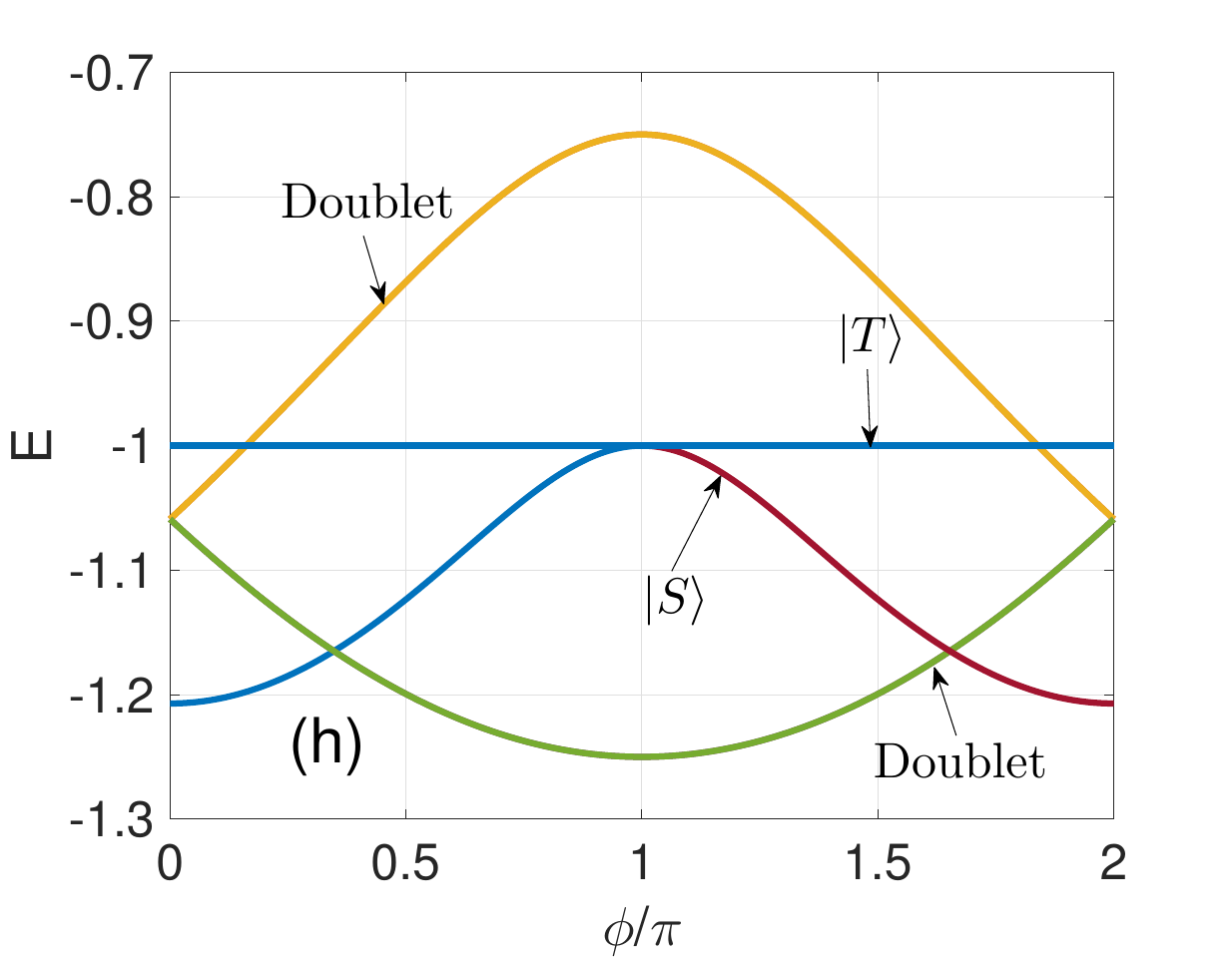}
		\end{minipage}
		\caption{(Color online) (a) Josephson current calculated by surrogate Hamiltonian and by low energy effective model.$\phi_{B}$ is taken to be $0.4\pi$ and $U_1=2,\varepsilon_1=-\frac{U_1}{2},\varepsilon_2=0,\Gamma=0.25$. (b) Entropy calculated by surrogate Hamiltonian. Parameters are set as Fig .4(a) .(c) Energy spectrum of the system, highlighting the lowest three eigenstates. The spectrum is calculate with low energy effective Hamiltonian. (d) Energy spectrum of the system, highlighting the lowest three eigenstates. The spectrum is calculate with surrogate Hamiltonian. Panels (e)--(h) correspond to (a)--(d), with the parameter $\phi_{B}=\pi$}
	\end{figure} 
	
	In Fig .4(a), we plot the Josephson current calculated using both the surrogate Hamiltonian and the low-energy effective Hamiltonian. The current exhibits a complex behavior and is not odd with respect to the superconducting phase difference $\phi$. In Fig .4(d), we show the spectrum obtained from the surrogate Hamiltonian, presenting only the lowest several levels. The figure reveals that the singlet-state energy level crosses the doublet-state level twice, which explains the unusual behavior: unlike the ordinary $0-\pi$ phase transition, the doublet state contributes both positive and negative components to the current. Fig .4(c) shows the spectrum from the low-energy effective Hamiltonian, which closely resembles that in Fig .4(d),demonstrating that the low-energy effective model successfully captures the key physical features of the system. Both the Josephson current and the phase transition points are in good agreement with the surrogate model results.
	The lowest several levels include the singlet, doublet, and triplet states, as indicated in Fig .4(c) and Fig .4(d). By varying the magnetic flux $\phi_{B}$ while keeping the interaction strength fixed, the phase transition points shift and reproduce the oddness with respect to the superconducting phase difference at $\phi_{B}=\pi$. The results are shown in Fig .4(e)–(h). 
	
	Comparing the two cases shown in Fig .4(d) and Fig .4(h), it is clear that the lowest several levels behave differently under varying magnetic flux. First, the triplet state never becomes the ground state in this setup. Since each quantum dot hosts one electron with spin up or spin down, the triplet state mediates single-particle transport. This process is strongly suppressed when $\phi_{B}$ are set to be $\pi$, as an electron passing through the upper dot interferes with one passing through the lower dot. This interference explains the nearly flat behavior of the $E(\phi)$ eigfunction. Similar to the SC–QD–SC case, single-particle transport contributes a negative current if the triplet were the ground state. Under arbitrary magnetic flux, the zero-crossing point of the current shifts only slightly, as the energy levels of QD1 and QD2 remain comparable.
	
    Second, the singlet state shows only a slight shift of its maximum point with changing magnetic flux. This occurs because two electrons, one spin-up and one spin-down, traverse the two quantum dots separately. Like the triplet case, this process is affected by the magnetic flux, causing the extreme points of the energy to shift correspondingly. The singlet state contributes a positive current, which is significantly larger than the current contribution from the triplet state, indicating subgap quasiparticle transport.
   
    The doublet state competes with the singlet state and drives the phase transition. Its phase is shifted almost linearly with $\phi_{B}$ which accounts for the phase transition point shifting. And it contributes the current both positive and negative depends on $\phi_{B}$. 
	
	Further, we directly diagonalized the low energy effective Hamiltonian to gain additional physical insight. Comparing Fig .4(c) and Fig .4(d), Fig .4(g) and Fig .4(h), triplet state energy level appears as a horizontal line in low-energy effective model, due to the omission of single particle excitation in the superconducting leads. Meanwhile singlet state anchors its maximum at phase difference $\pi$. By inspecting the relatively small $16\times 16$ Hilbert space and performing appropriate unitary transformations, we find that the singlet state couples only to the other two states $\left(|0\rangle+|d\rangle\right)_1\otimes\left(|0\rangle-|d\rangle\right)_2$ and $\left(|0\rangle-|d\rangle\right)_1\otimes\left(|0\rangle+|d\rangle\right)_2$ in  low energy effective model when $\varepsilon=-\frac{U}{2}$. This coupling is captured by the following $ 3\times 3$matrix:
	\begin{equation}
		\begin{bmatrix}
			2\!\sin(\frac{\phi}{2})\!\sin(\frac{\phi_{B}}{2})\Gamma&0&-\!\sqrt{2}\cos(\frac{\phi}{2})\Gamma\\
			0&-2\!\sin(\frac{\phi}{2})\!\sin(\frac{\phi_{B}}{2})\Gamma&-\!\sqrt{2}\cos(\frac{\phi}{2})\Gamma\\
			-\!\sqrt{2}\cos(\frac{\phi}{2})\Gamma&-\!\sqrt{2}\cos(\frac{\phi}{2})\Gamma&\varepsilon_1
		\end{bmatrix}.
	\end{equation}
	The coupling strength is $\sqrt{2}\cos(\frac{\phi}{2})\Gamma$, which is $\sqrt{2}$ times cross dot paring potential $\cos(\frac{\phi}{2})\Gamma$ and is independent of $\phi_{B}$. Consequently, the maximum of $|S\rangle$ energy is fixed. The eigenvalues of this $3\times 3$ matrix can be solved analytically using Cardano's formula. 
	
	On the other hand, after unitary transforming, the doublet state stem from the coupling of state $|d\!\uparrow\rangle+|0\!\uparrow\rangle$,  $|\uparrow\! d\rangle+|\uparrow\! 0\rangle$ and $|d\!\uparrow\rangle-|0\!\uparrow\rangle$,  $|\uparrow\! d\rangle-|\uparrow\! 0\rangle$ with coupling matrix as follows at $\varepsilon_1=-\frac{U_1}{2}$:
	\begin{equation}
		\begin{bmatrix}
			H_1&0\\
			0&H_2
		\end{bmatrix},
	\end{equation}
	where:
	\begin{equation}
		\begin{split}
			&H_1=\begin{bmatrix}
				\Gamma\!\cos(\frac{\phi-\!\phi_{B}}{2})&-\Gamma\!\cos(\frac{\phi}{2})\\
				-\Gamma\!\cos(\frac{\phi}{2})&-\!\frac{U_1}{2}\!+\!\Gamma\!\cos(\frac{\phi+\!\phi_{B}}{2})
			\end{bmatrix},\\
			&H_2=\begin{bmatrix}
				-\Gamma\!\cos(\frac{\phi-\!\phi_{B}}{2})&\Gamma\!\cos(\frac{\phi}{2})\\
				\Gamma\!\cos(\frac{\phi}{2})&-\!\frac{U_1}{2}\!-\!\Gamma\!\cos(\frac{\phi+\!\phi_{B}}{2})
			\end{bmatrix}.
		\end{split}
	\end{equation}
	The eigvalues are roots of these two quadratic equations. The lowest two contribute the current with one stem from the coupling of  $|d\!\uparrow\rangle-|0\!\uparrow\rangle$,  $|\uparrow\! d\rangle-|\uparrow\! 0\rangle$ generates a positive current, while the one arising from the coupling of $|d\!\uparrow\rangle+|0\!\uparrow\rangle$,  $|\uparrow\! d\rangle+|\uparrow\! 0\rangle$ generates a negative current at $\phi_{B}=0.4\pi$, as shown in Fig .4(a) and Fig .4(c).
	
	For a more general case that $\varepsilon_1\neq-\frac{U_1}{2}$, the four states mentioned above couple with each other, resulting in a quartic equation. Unlike the previous quadratic equations, which involve Cooper-pair transport separately through the two quantum dots, the quartic equation also includes local Cooper-pair transport. This process is reflected in the coupling matrix by the interaction between  $|d\!\uparrow\rangle-|0\!\uparrow\rangle$ and $|d\!\uparrow\rangle+|0\!\uparrow\rangle$,with a coupling factor $\epsilon=\varepsilon_1+\frac{U_1}{2}$. Meanwhile, Meanwhile, the cubic equation determining the eigenvalue of the $|S\rangle$ state now becomes quintic, which can only be solved numerically.
	
	\subsection{$U_1=U_2\neq 0$}
	For the case that both two QDs include Coulomb interaction, Ruderman-Kittel-Kasuya-Yosida(RKKY) interaction starts to play a significant role\cite{PhysRevB.74.045312},inducing a phase transition between the $|S\rangle$ and $|T\rangle$ states and leading to ferromagnetic spin correlations. The results are shown in Fig .5(a). For relatively weak interaction strength, the doublet state remains relevant, so the system undergoes two successive phase transitions: first from the doublet state to the singlet state, and then from the singlet state to the triplet state, as shown in Fig .5(b). Both the doublet state and the $|S\rangle$ state contribute positively to the Josephson current, while the $|T\rangle$ state contributes negatively. This is because the $|T\rangle$ current originates from single-particle transport, whose energy lies above the superconducting gap, resulting in a consistently negative contribution.
	\begin{figure}[htbp]
		\centering
		\includegraphics[width=0.91\linewidth,margin={0cm 0cm 0.3cm 0cm}]{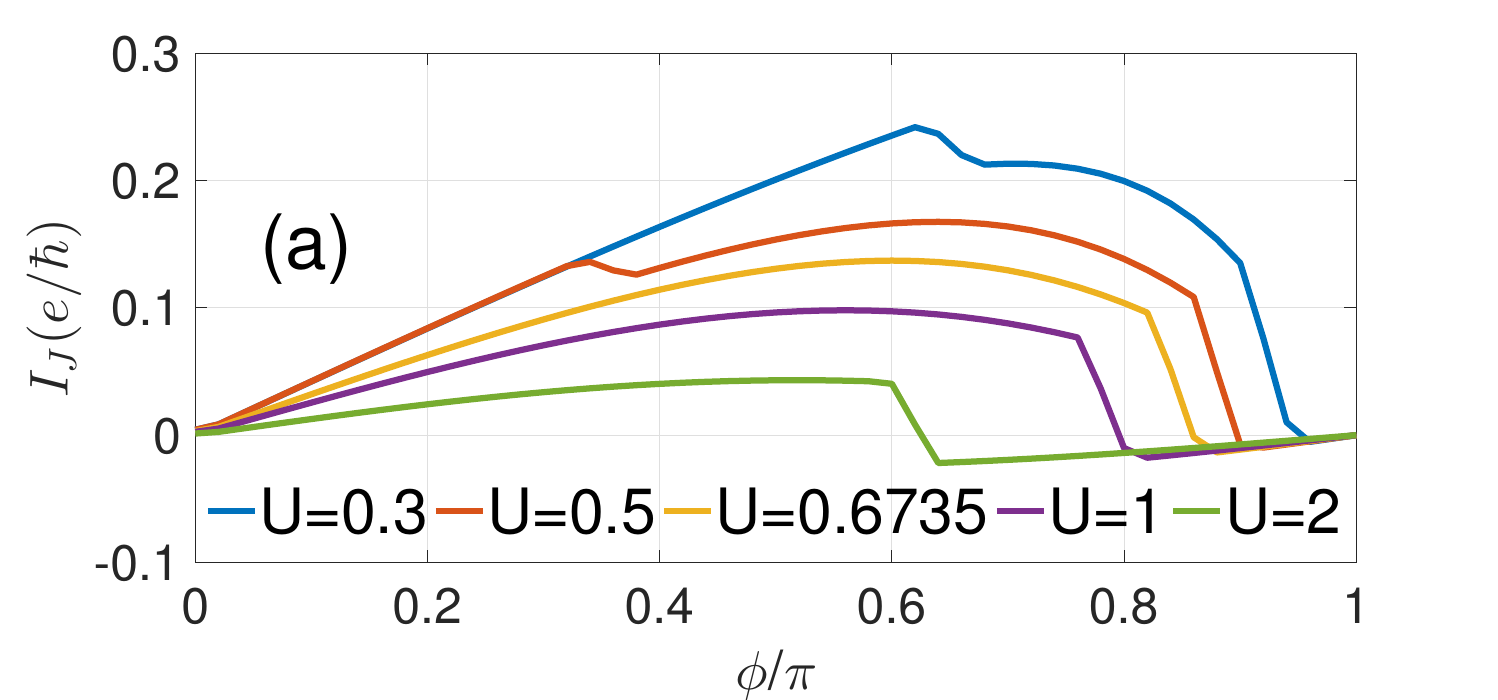}
		\includegraphics[width=0.95\linewidth]{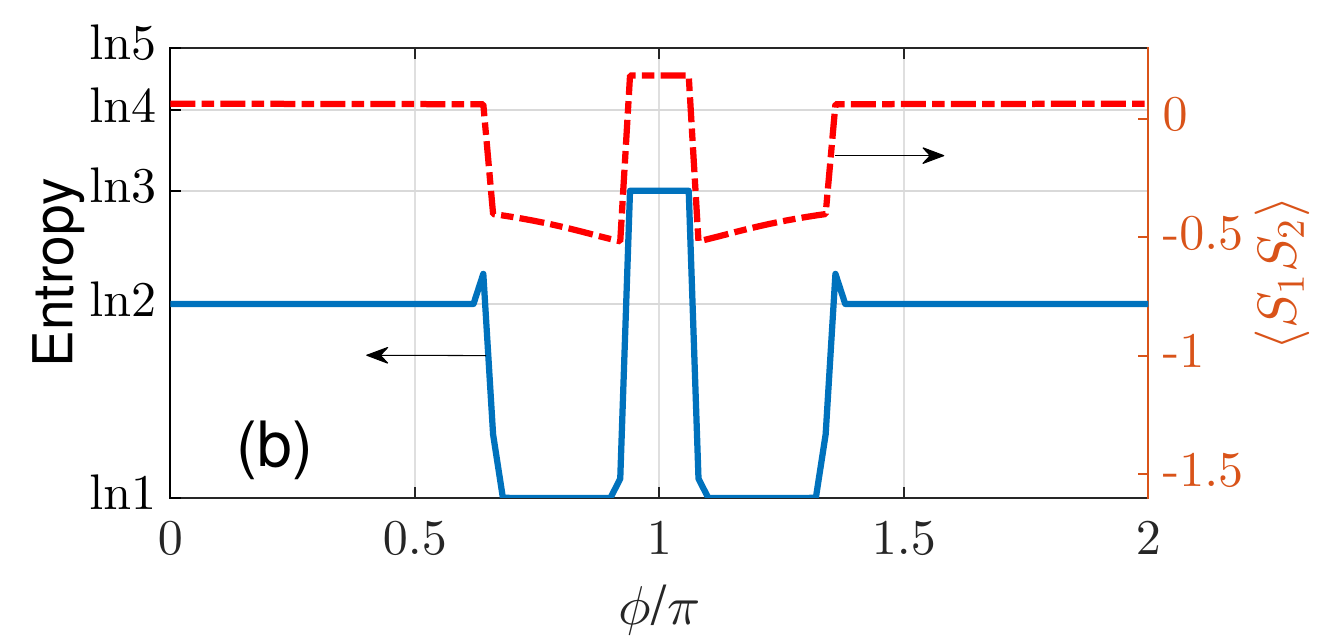}
		\caption{(Color online) (a) Current phase relationship for different U. Parameters are set as follows: $\phi_{B}=0$ and $\Gamma=0.25$. (b)Entropy with $\phi$ at $U_1=U_2=0.3$.}
	\end{figure} 
	As the interaction strength increases, transport through the doublet state is completely overwhelmed by the single-particle transport of the $|T\rangle$ state, as shown in Fig .5(c). In this regime, only a single phase transition occurs between the $|S\rangle$ and $|T\rangle$ states. Eventually, for large 
	U, the Josephson current becomes entirely a $\pi$-junction current. Interestingly, the current contribution from the doublet state remains unaffected by U. In the opposite limit, as $U_1=U_2\to 0$, the doublet state dominates. However, the triplet state remains near the phase difference $\phi=\pi$ until $U_1=U_2=0$.
	
	In the presence of $\phi_{B}$, transport through both the doublet and triplet states is suppressed. The corresponding phase diagrams are presented in Fig .6(a) and Fig .6(b). Blue denotes $|S\rangle$ phase region, yellow denotes $|T\rangle$ phase region and green represents doublet region. To interpret these results, we once again employ the low-energy effective Hamiltonian. Although, in the case where both sites host Coulomb interactions, the low-energy effective model does not reproduce the surrogate Hamiltonian results with high accuracy, it nevertheless provides valuable physical insight into the underlying mechanisms.                                                         
	\begin{figure}[htbp]
		\centering
		\includegraphics[width=0.75\linewidth]{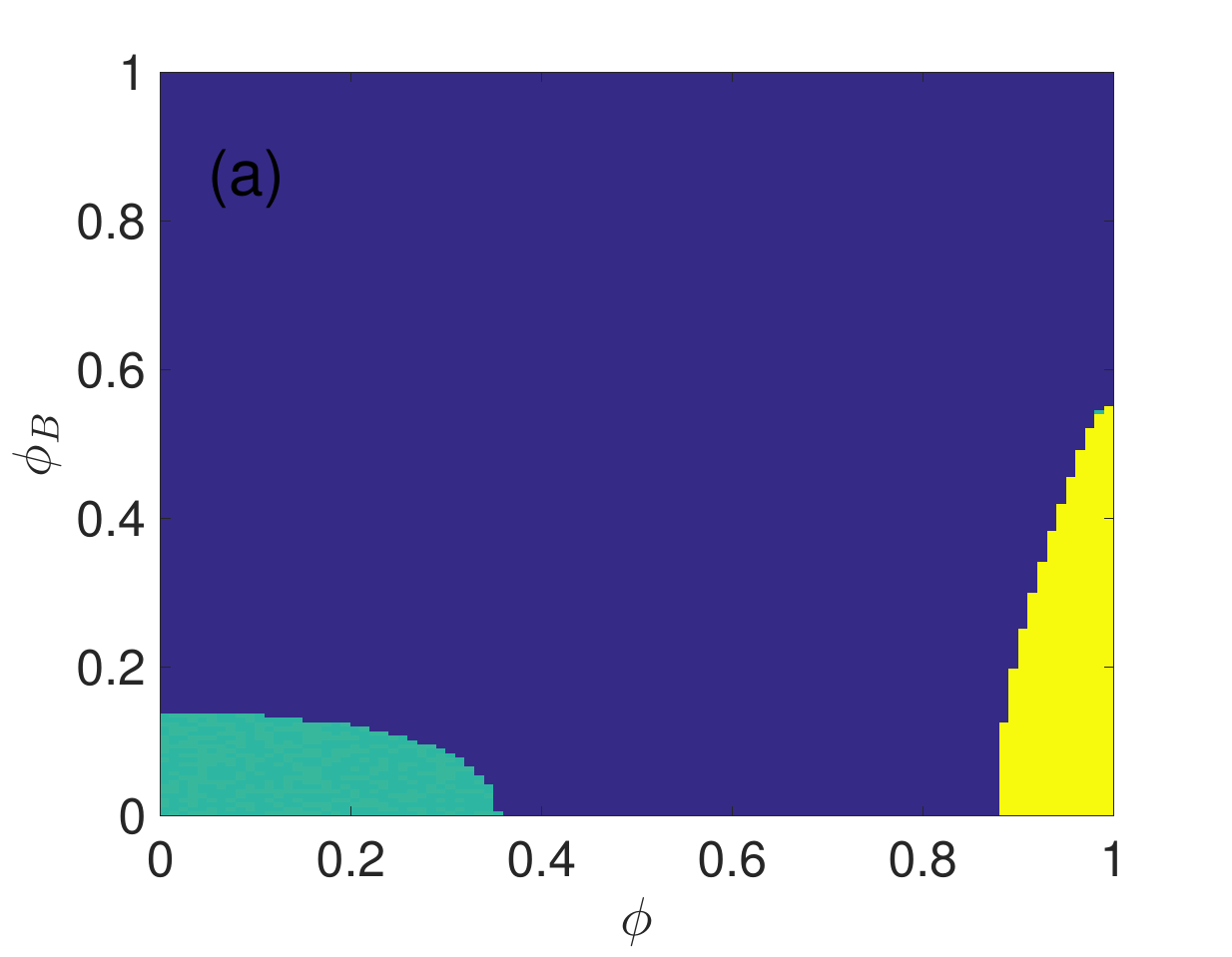}
		\includegraphics[width=0.75\linewidth]{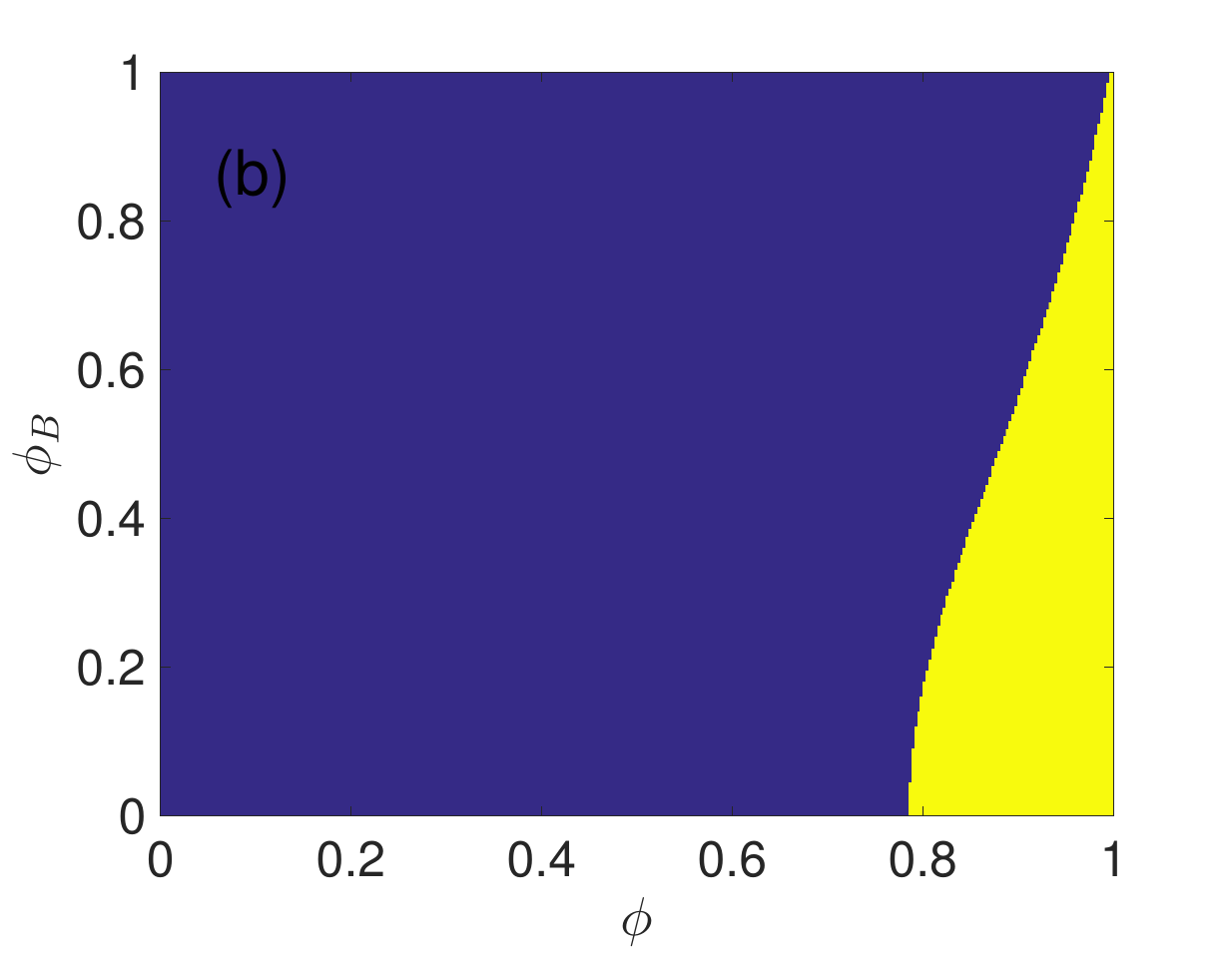}
		\caption{(Color online) Phase diagram versus $\phi$ and $\phi_{B}$ (a) $U_1=U_2=0.5$, $\Gamma=0.25$ (b)$U_1=U_2=1$, $\Gamma=0.25$. }
	\end{figure} 
	
	Unlike the setup discussed in Section B, in the symmetric QD configuration the doublet state does not shift its extremal value with $\phi_{B}$.  Within the low-energy effective model, the relevant subspace is spanned by $|d_1\!\uparrow_2\rangle,|\uparrow_1 \! d_2\rangle,|0_1\!\uparrow_2\rangle,|\uparrow_1\!0_2\rangle$ , and the corresponding coupling matrix is given by:
	\begin{equation}
		\begin{bmatrix}
			-\frac{U_2}{2}&0&\!\Gamma\!\cos(\frac{\phi-\!\phi_{B}}{2}\!)&-\!\Gamma\!\cos(\frac{\phi}{2})\\
			0&-\frac{U_1}{2}&-\!\Gamma\!\cos(\frac{\phi}{2})&\!\Gamma\!\cos(\frac{\phi+\!\phi_{B}}{2}\!)\\
			\!\Gamma\!\cos(\frac{\phi-\!\phi_{B}}{2}\!)&-\!\Gamma\!\cos(\frac{\phi}{2})&-\frac{U_2}{2}&0\\
			-\!\Gamma\!\cos(\frac{\phi}{2})&\!\Gamma\!\cos(\frac{\phi+\!\phi_{B}}{2}\!)&0&-\frac{U_1}{2}
		\end{bmatrix}.
	\end{equation}
	Thus, the transport process involves both local and non-local Cooper pair tunneling, in contrast to the single-site-interaction case where only the non-local channel contributes. The corresponding characteristic equation can be solved analytically, yielding:\par
	\begin{widetext}
		\vspace{-0.5\baselineskip}  % 增加上下間距避免擠壓
		\begin{equation}
			E_{Doublet}=\frac{1}{2} \left(\pm\sqrt{\Gamma ^2 ((\cos (\phi )-1) \cos (\phi_B)+\cos (\phi )+3)}\pm 2 \Gamma
			\cos \left(\frac{\phi }{2}\right) \cos \left(\frac{\phi_B}{2}\right)-U\right).
		\end{equation}
		\vspace{-0.5\baselineskip}  % 增加上下間距避免擠壓
	\end{widetext}
	
	The solution is a monotonic increasing(decreasing) function of $\phi$ and cross each other at $\phi=\pi$ point taking the value $\frac{1}{2} \left(-\sqrt{\Gamma ^2 (2-2 \cos (\phi _B))}-U\right)$. Moreover, the energy difference $\Delta E(\phi_{B})=\left.E_{Doublet}\right|_{\phi = \pi}-\left.E_{Doublet}\right|_{\phi = 0}$ vanishes at $\phi_{B}=\pi$, indicating that the local and non-local transport channels cancel each other at this point, leading to a zero Josephson current. This explains both the emergence of the doublet phase region near $\phi=0$ and the shrinking of this region as $\phi_{B}$ increases from 0 to $\pi$.
	
	for $|T\rangle$ state, previous studies using fourth-order perturbation theory \cite{PhysRevB.74.045312} \cite{PhysRevB.94.155445} indicated that the  energy correction of co-tunneling spin exchange process may let $|T\rangle$ be the ground state and the ratio between
	triplet-favoring and singlet-favoring is $1+\varepsilon_{DQD}/\Delta$ where $\varepsilon_{DQD}$ is QDs excitation energy and in our case is $\frac{U}{2}$. This explains the triplet-favoring with increasing U. And by letting $\phi_{B}=\pi$, this co-tunneling process interfere destructively and only Cooper pair tunneling process maintains which made the ground state keep singlet. 
	The $|S\rangle$ in this case in low energy effective model is coupled with the other four states by the coupling matrix as:
	\begin{widetext}
		\vspace{0.1\baselineskip}  
		\begin{equation}
			\begin{bmatrix}
				0&\!\Gamma\!\cos(\frac{\phi-\!\phi_{B}}{2}\!)&\!\Gamma\!\cos(\frac{\phi+\!\phi_{B}}{2}\!)&0&-\sqrt{2}\Gamma\!\cos(\frac{\phi}{2})\\
				\!\Gamma\!\cos(\frac{\phi-\!\phi_{B}}{2}\!)&0&0&\!\Gamma\!\cos(\frac{\phi+\!\phi_{B}}{2}\!)&0\\
				\!\Gamma\!\cos(\frac{\phi+\!\phi_{B}}{2}\!)&0&0&\!\Gamma\!\cos(\frac{\phi-\!\phi_{B}}{2}\!)&0\\
				0&\!\Gamma\!\cos(\frac{\phi+\!\phi_{B}}{2}\!)&\!\Gamma\!\cos(\frac{\phi-\!\phi_{B}}{2}\!)&0 &\sqrt{2}\Gamma\!\cos(\frac{\phi}{2})\\
				-\sqrt{2}\Gamma\!\cos(\frac{\phi}{2})&0&0&\sqrt{2}\Gamma\!\cos(\frac{\phi}{2})&-U
			\end{bmatrix}.
		\end{equation}
		\vspace{0.1\baselineskip} 
	\end{widetext}
	
	The relevant subspace consists of $|0_1,0_2\rangle$, $|d_1,0_2\rangle$, $|0_1,d_2\rangle$, $|d_1,d_2\rangle$, $|S\rangle$. In this framework, tunneling occurs between $|S\rangle$ and another four mixture and is irrelevant with magnetic flux which alternates the coupling strength between these four states. As a result, the Josephson current hold non-zero at $\phi_{B}=\pi$. Moreover, the maximum of this non-zero current, as critical current, displays a pronounced peak with a sudden rise followed by a rapid fall with increasing U. The result are shown in Fig .7. 
	\begin{figure}[htbp]
		\centering
		
		\begin{minipage}{0.79\linewidth}
			\includegraphics[width=1.05\linewidth]{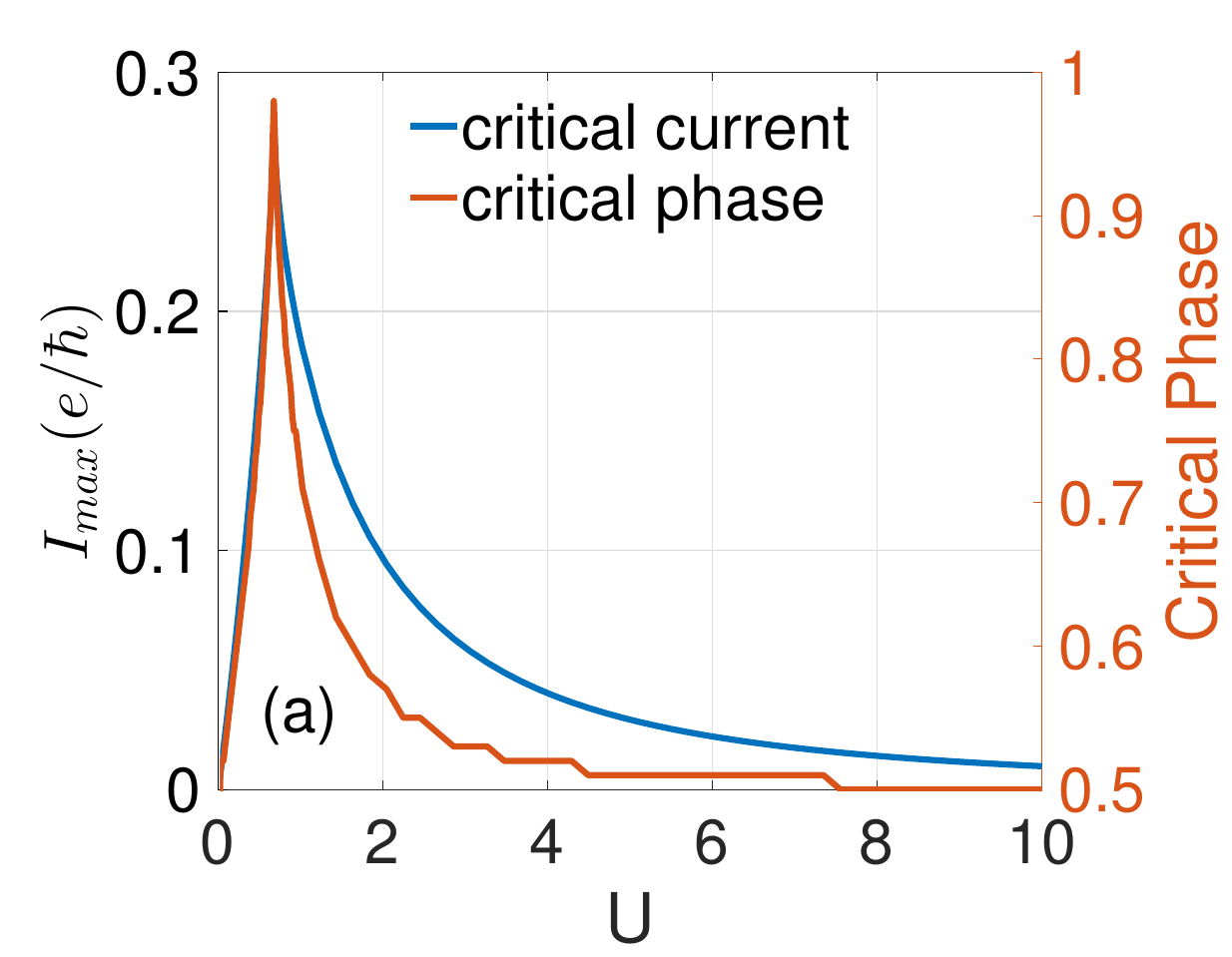}
			\includegraphics[width=0.95\linewidth]{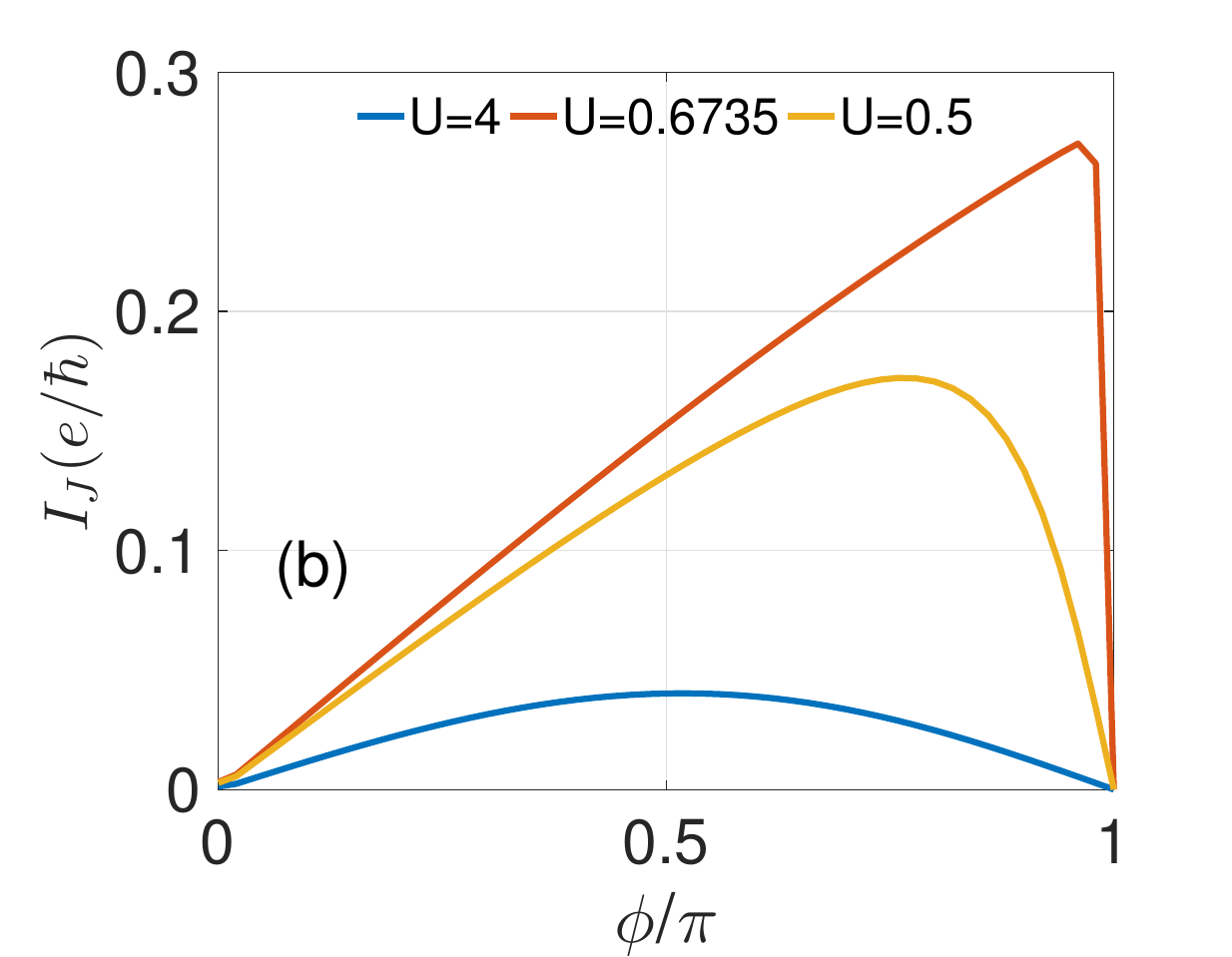}
		\end{minipage}
		\caption{(Color online) (a) Critical current and critical phase versus U, $\Gamma=0.25$ (b)Josephson Current at $U=4$, $U=0.6735$ and $U=0.5$}
	\end{figure}
	Here we fix $\varepsilon_d$ at the particle–hole symmetric point. As shown in Fig .7(a), the corresponding critical phase is pinned at $\phi=\pi$ when the critical current reaches its maximum at $U=U^\star\approx0.6735$. In contrast, in the limits $U \!\to\! 0$ and $U \!\to\!\infty$, critical phase is pinned at  $\phi=\frac{\pi}{2}$, which with a sine-like current profile. Remarkably, at $U=U^\star$, Josephson current is like the normal 0 phase current in S-QD-S junction which grows almost linearly with SC phase difference $\phi$, as illustrated in Fig .7(b). Throughout, the ground state remains $|S\rangle$ and no phase transition occurs, in contrast to the conclusion of Ref.\cite{PhysRevB.94.155445}, where the maximum critical current was found at the phase boundary.
	
   Inspection of the low-energy effective Hamiltonian nevertheless reveals the existence of a maximum critical current as $U$ is varied, but at a distinct value $U^\star_{\mathrm{eff}}=0.5$, different from that obtained from the surrogate Hamiltonian. At this special interaction strength, the energy levels of the doublet and triplet states coincide. Since $|S\rangle$ and $|T\rangle$ are degenerate at $\phi=\pi$, this coincidence causes $|S\rangle$ to align with the doublet state at $\phi=\pi$, thereby facilitating tunneling, enhancing the critical current, and shifting the critical phase to $\pi$.
   
   For surrogate Hamiltonian, doublet and triplet state energy level cross at $\phi=\pi$ when $U=0.6735$. And there is a gap between the $|S\rangle$ and doublet state energy level at $\phi=\pi$ if $U<U^\star$. While for $U>U^\star$, this gap disappear but the second ground state is now to be $|T\rangle$ other than the doublet state. So there is a phase transition between the upper $|S\rangle$ and lower $|S\rangle$ which one conicides with $|T\rangle$ at $\phi=\pi$ and another has a gap with the doublet state. This fact indicates that although the ground state is held to be $|S\rangle$ in the presence of magnetic flux, it still can has a inner phase transition of  $|S\rangle$. This also accounts for the special Josephson current profile that when $U=U^\star$, two subgap Andreev bound state merge in one which is just the situation of 0 phase S-QD-S junction. 
	
	In a word, in the presence of magnetic flux, local tunneling is inhibited and the phase transition with increasing U is no more $0-\pi$ transition between $|S\rangle$ and $|T\rangle$ but $0-0$ transition between upper $|S\rangle$ and lower $|S\rangle$. The Josephson current is none-zero although $\phi_{B}=\pi$.
	\subsection{$\Gamma > \Delta$}
	Here we set $\Gamma>\Delta$ which is quit away from the perturbation region that is suitable for Schrieffer-Wolff transformation. Also the low energy effective model did not perform well since now the quasi-particle excitation is not negligible. So we just focus on the numerical result and give some qualitative explanation.
	
	By increasing $\Gamma$, the doublet phase region can now touch with triplet phase region. And in the presence of magnetic flux, these two phases will eventually turn into singlet as it is in the section above. So there should be a triple point as it is shown in Fig .8(a). As we can see that with large $\Gamma$, the single particle tunneling became important and the doublet state phase area inflates. And like the condition in previous section, the doublet and triplet energy level can meet at $\phi=\pi$. But it is different that now the state sequence(from button to above) at $\phi=\pi$ point is no longer $|S\rangle\to doublet\to |T\rangle\to |S\rangle$ but $|T\rangle\to doublet\to |S\rangle\to |S\rangle$. It is because when single particle tunneling dominates, it is prefer $|T\rangle$ and doublet state than Cooper pair tunneling related $|S\rangle$. And the magnetic flux more strongly suppressed $|T\rangle$ than the doublet which made a 'phase peninsula' of the doublet state. 
	
	Again there exists the $|S\rangle\to|S\rangle$ inner phase transition when $U^\star=10.83$ with $\phi_{B}=\pi$. Which means although it is not well gapped when $\Gamma>\Delta$, magnetic flux can neutralize this and induce a similar behaviour like it is well gapped by $\Delta$.
	\begin{figure}[htbp]
		\centering
		\begin{minipage}{0.79\linewidth}
			\includegraphics[width=0.95\linewidth]{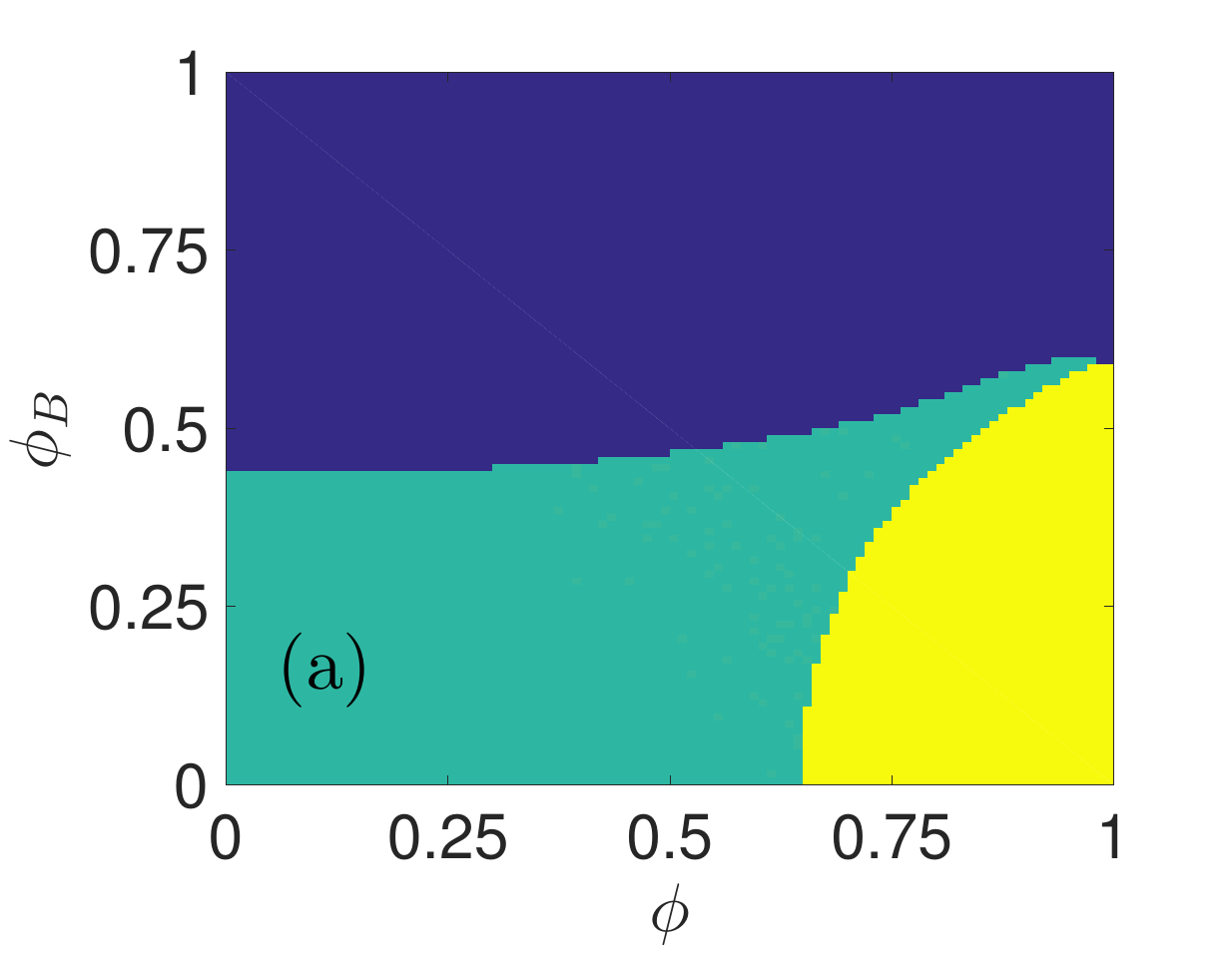}
			\includegraphics[width=0.95\linewidth]{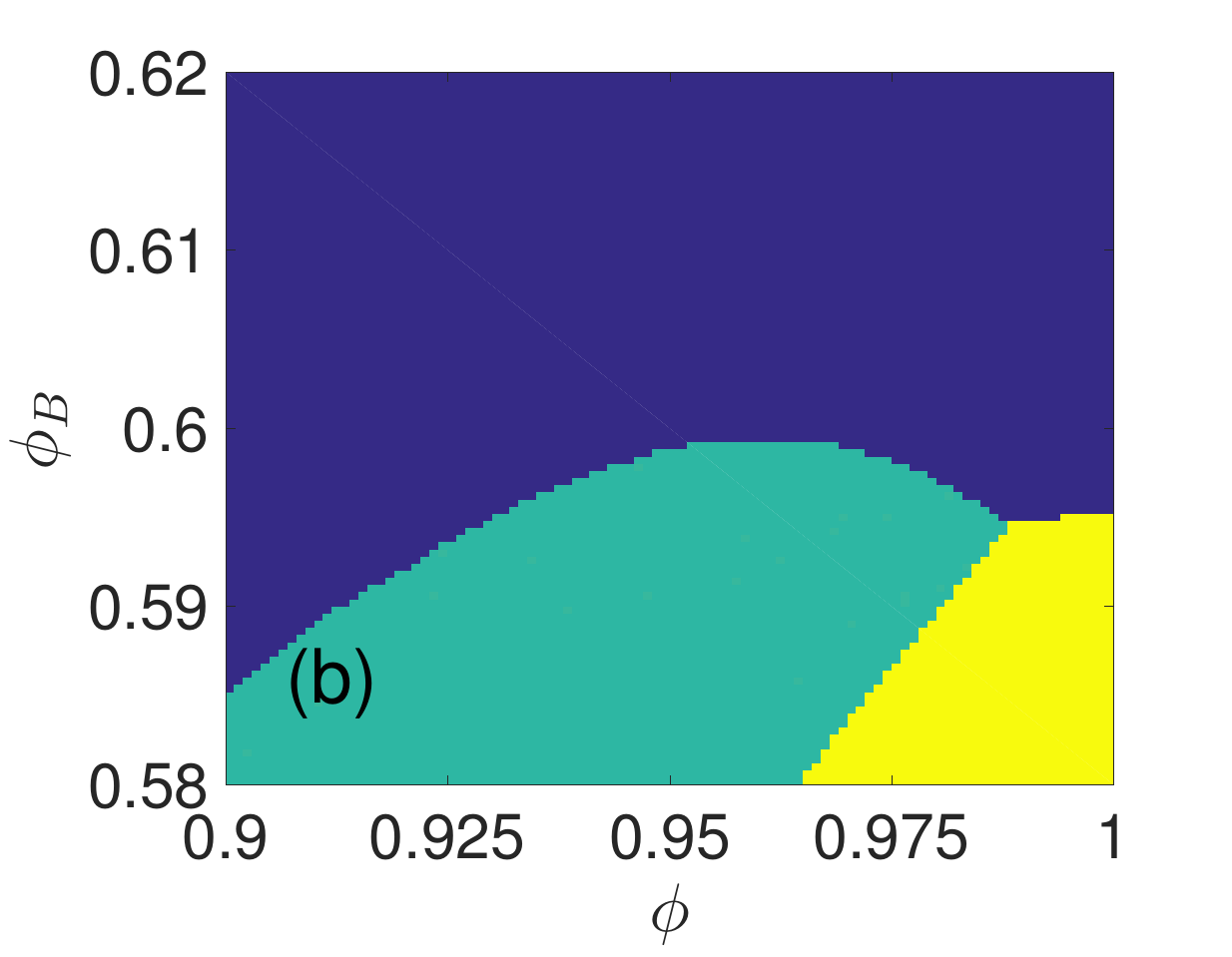}
		\end{minipage}
		\caption{(Color online) Phase diagram versus $\phi$ and $\phi_{B}$ (a) $U_1=U_2=8$, $\Gamma=2$.(b)Enlarged view of the triple point region.}
	\end{figure}
	\section{Conclusion}
	In this study, we investigated a Josephson junction consisting of a QD parallel connected to another one and attach to two superconducting leads. By employing a discretization approach, we discretized the self-energy of the superconducting leads, thereby replacing the Hamiltonian of the leads with that of a finite number of discrete points. Furthermore, we performed exact diagonalization of the system's effective Hamiltonian using the state-space expansion method. Subsequently, we numerically computed the system's entropy, parity, spin correlation between the two QDs and the Josephson current under various conditions.  Meanwhile, a low effective Hamiltonian is employed to account for the numeric results and get some physical insight. When the Coulomb interaction is absence, imagery time path integral approach is used to make a precise calculation of the current. This result is compared with the current calculated from surrogate model. It is shown that these two methods reach a very well agreement.
	
	we found when none of the dot include Coulomb interaction and $\varepsilon_d$ is set symmetry, the Josephson current vanishes if $\phi_{B}=\pi$ due to the destructive interference. By setting on site potential asymmetry, a $0-\pi$ phase transition can be achieved by increasing $\phi_{B}$ while the ground state is held to be $|S\rangle$. The phase transition is mainly derive from the changing of subgap discrete states contribution. The overgap continuum contribution changes little with the varying magnetic flux.
	
	When only one dot holds the Coulomb interaction, there will be a magnetic flux controlled phase transition between doublet and singlet state.  The phase changing point is shifted almost linearly with $\phi_{B}$. With increasing U, the $|S\rangle$ phase area shrinks. In this case, $|T\rangle$ didn't participate in the phase changing and the low energy effective model work well and can give a closed result with the surrogate model.
	
	At the case two dots both hold Coulomb interaction, $|T\rangle$ has the opportunity to be the ground state. For $\Gamma<\Delta$, if U is small, the Josephson current can experience phase transition twice versus $\phi$. But for large U, doublet state phase area disappear and only $|S\rangle\to|T\rangle$ phase transition remains. Presence of magnetic flux inhibits these two phase and when $\phi_{B}=\pi$, it is all singlet ground state.
	
	Contrary to the interaction absence case, Josephson current is now non-zero when $\phi_{B}=\pi$. A inner phase transition with upper $|S\rangle$ and lower $|S\rangle$ happens which cause a critical current peak with U increase. 
	
	For $\Gamma>\Delta$, doublet state phase area inflates and tough the $|T\rangle$ phase area. By increasing $\phi_{B}$, these two phase turn into $|S\rangle$ and there so exists a triple point.
	\bibliography{ref_v2}
\end{document}